\documentclass[aps, prl,superscriptaddress, groupedaddress, floatfix, reprint]{revtex4-2} 

\usepackage{graphicx}
\usepackage{dcolumn}
\usepackage{bm}
\usepackage{amsfonts}
\usepackage{amsmath}
\usepackage{amssymb}
\usepackage{color}
\usepackage[normalem]{ulem}
\usepackage[percent]{overpic}
\usepackage{natbib}
\usepackage{soul}

\usepackage[colorlinks=true,citecolor=blue]{hyperref}
\hypersetup{colorlinks=true,citecolor=blue,linkcolor=red,urlcolor=blue}

\def\be{\begin{equation}}
\def\ee{\end{equation}}

\def\kv{{\bf k}}

\begin{document}

\title{Engineering the Bogoliubov Modes through Geometry and Interaction: From Collective Edge Modes to Flat-band Excitations}

\author{Maryam Darvishi}
	\affiliation{Department of Physics, Institute for Advanced Studies in Basic Sciences (IASBS), Zanjan 45137-66731, Iran}
\author{Fatemeh Pouresmaeeli}
\affiliation{Department of Physics, Institute for Advanced Studies in Basic Sciences (IASBS), Zanjan 45137-66731, Iran}
\author{Saeed H. Abedinpour}
\email{abedinpour@iasbs.ac.ir}
	\affiliation{Department of Physics, Institute for Advanced Studies in Basic Sciences (IASBS), Zanjan 45137-66731, Iran}

\date{\today}

\begin{abstract}
	We propose a procedure to engineer solid-state lattice models with superlattices of interaction-coupled Bose-Einstein condensates. We show that the dynamical equation for the excitations of Bose-Einstein condensates at zero temperature can be expressed in an eigenvalue form that resembles the time-independent Schr{\"o}dinger equation. The eigenvalues and eigenvectors of this equation correspond to the dispersions of the collective modes and the amplitudes of the density oscillations. 	
	This alikeness opens the way for the simulation of different tight-binding models with arrays of condensates. We demonstrate, in particular, how we can model a one-dimensional Su-Schrieffer-Heeger lattice supporting topological edge modes and a two-dimensional Lieb lattice with flat-band excitations with superlattices of Bose-Einstein condensates. 	
\end{abstract}

\maketitle

\emph{Introduction.}--
The quest for artificial models that can simulate the behavior of solid-state electronic systems is a fundamentally and practically exciting and fruitful endeavor. 
Two major areas in condensed matter physics that have attracted enormous interest lately are the topologically nontrivial behavior of materials~\cite{Asbth_Book2016,Moessner2021} and the flat-band electronic dispersions responsible for the strongly correlated phases of materials~\cite{Regnault2022}. Much work has also been done to realize both phenomena in non-solid-state systems~\cite{Krishnamoorthy2023,Leykam2018}. 
The topological properties of materials usually refer to their single-particle electronic structures in which the topological indices are defined in terms of the electronic Bloch states~\cite{Moessner2021}. In recent years, the same concept has been extended to other platforms such as mechanical systems~\cite{ma2019topological,manda2022nonlinear}, photonic crystals~\cite{Haldane_PRL2008,Raghu_PRA2008,jiang2021topological,cheng2022topological}, and electronic circuits~\cite{imhof2018topolectrical,liu2022fully}. 

Collective density oscillations in many-body systems originate from inter-particle interactions. As a paramount example, plasmons emerge due to the long-range Coulomb interaction between charged particles~\cite{giuliani2005quantum}. 
An interesting question is whether these collective excitations could also exhibit topological behaviors. 
Even though there are examples where the collective excitations of topologically nontrivial electronic systems are also localized~\cite{guan2022plasmons}, our main quest here is to investigate if it would be possible to have localized edge collective excitation in systems whose single particle excitations are topologically trivial. Such a topological excitation, in principle, may arise from tuning the particle-particle interaction. 
Furthermore, we are also interested in exploring under what circumstances it would be possible to realize flat-band (i.e., dispersion-less) collective excitations.
We should note that even though flat-bands and topological Bogoliubov modes have been studied in bosonic systems~\cite{Jalali-mola_PRL2023,Tesfaye_arxiv2024}, these behaviors mainly originate from the hopping of bosons in an optical lattice~\cite{Lewenstein_Book2016,Pitaevskii_Book2016}, rather than the inter-particle interactions.

In this letter, we start with the linear response equation for the density oscillations of an array of Bose-Einstein condensates (BECs). 
With some straightforward rearrangements, we transform the linear response equation in the resonance condition into a standard eigenvalue problem. This eigenvalue equation, whose eigenvalues are the square modulus of the excitation spectrums, looks similar to the time-independent Schr{\"o}dinger equation. Therefore, the presumed topological behavior could be engineered through the geometric phase of the eigenvectors, which in this case are simply the vector of the density oscillations in different condensates.    

We showcase the topological collective excitation in a one-dimensional bipartite superlattice of BECs. This simple structure is equivalent to the renowned Su-Schrieffer-Heeger (SSH) lattice~\cite{Moessner2021,Su_PRL1979}, one of the simplest topologically nontrivial electronic models. For suitable choices of the inter-cell and intra-cell interactions, this SSH BEC array supports collective excitations localized at the edges of the superlattice.
To realize flat-band collective excitations, we propose a two-dimensional superlattice of the so-called Lieb lattice, consisting of quasi-one-dimensional tubes of BECs.  Such a structure can support a dispersion-less collective density oscillation mode.

\emph{Model.}--
We consider an array of $N$ Bose-Einstein condensates. 
Each condensate consists of particles with mass $m_i$ and average particle density of $n_i$. 
Such a setup could be readily realized experimentally with ultra-cold atoms trapped in optical lattices~\cite{Lewenstein_Book2016}.
An applied external potential $V_j^{ext}(q,\omega )$, which couples to the density of particles on the $j^{\rm th}$ condensate, induces density fluctuations of wave-vector $q$ and frequency $\omega$, as~\cite{giuliani2005quantum}
\be\label{eq:density_osc}
\delta {n_i}(q,\omega ) = \sum\limits_{j = 1}^N {{\chi _{ij}}(q,\omega )V_j^{ext}(q,\omega )},
\ee
where $i = 1,...,N$, and $\chi(q,\omega)$ is the interacting linear density-density
response function matrix, which in the random-phase approximation (RPA) reads
\be
{\chi} (q,\omega ) = {\left[ {\mathcal I - \Pi (q,\omega ) V(q)} \right]^{ - 1}} \Pi (q,\omega ).
\ee
Here, $\mathcal I$ is a $N \times N$ unit matrix, $\Pi (q,\omega )$ is the matrix of non-interacting density-density
response functions, and the interaction matrix elements, i.e., $V_{ij}(q)$, refer to the interaction between two bosons from condensates $i$ and $j$. The dispersion of collective modes is obtained from the resonance condition for Eq.~\eqref{eq:density_osc}
\be\label{eq:resonance}
{\chi ^{ - 1}}(q,\omega )\delta n(q,\omega ) = 0,
\ee
where ${\chi ^{ - 1}}(q,\omega )={\Pi ^{ - 1}}(q,\omega )-V(q )$. 
We should note that what the RPA gives for the collective modes' dispersions is identical to the mean-field Bogoliubov-de Gennes (BdG) excitation spectrum at the vanishing temperature~\cite{tosi2006manybody}.  
If there is no tunneling between different BECs, the non-interacting density-density response matrix is diagonal and at zero-temperature reads~\cite{Pitaevskii_Book2016,tosi2006manybody}
\be
\Pi _{ij}(q,\omega ) = \delta_{i,j}\frac{{2{n_i}{\varepsilon _{q,i}}}}{{{{(\hbar \omega )}^2} - \varepsilon _{q,i}^2}}.
\ee
Here, ${\varepsilon _{q,i}} = {\hbar ^2}{q^2}/(2{m_i})$ is the non-interacting dispersion of a particle at the $i^{\rm th}$ condensate. 
With some straightforward rearrangements, we can rewrite Eq.~\eqref{eq:resonance} in a generalized eigenvalue problem form
\be\label{eq:gen_eigen}
 D(q)\delta n(q,E ) = {E^2}  S(q)\delta n(q,E ),
\ee
where $E\equiv \hbar \omega$,  and we have introduced
${D}_{i j}(q)\equiv V_{ij}(q)+\delta_{i, j} \varepsilon_{q, i}/(2 n_{i})$
and
${S}_{ij}(q)\equiv {\delta_{i, j}}/{(2 n_{i} \varepsilon_{q, i})}$.
As the ${S}$-matrix is diagonal, Eq.~\eqref{eq:gen_eigen} could be rewritten in a standard eigenvalue problem form
\be\label{eq:eigen}
{\mathcal{D}}(q) \delta \widetilde {n}(q, E)=E^{2} \delta \widetilde {n}(q, E),
\ee
with ${\mathcal{D}}(q)\equiv {S}^{-\frac{1}{2}}(q) {D}(q){S}^{-\frac{1}{2}}(q)$ and $\delta\widetilde {n}(q, E)\equiv {S}^{\frac{1}{2}}(q)\delta n(q, E)$. 
The dispersions of the collective modes $E_\lambda(q)$ (with $\lambda=1,\cdots,N$), are obtained from the eigenvalues of Eq.~\eqref{eq:eigen}, and the corresponding density oscillations from its eigenvectors $\delta n_\lambda(q,E)= {S}^{-\frac{1}{2}}(q)\delta\widetilde {n}_\lambda(q, E)$.

Note that Eqs.~\eqref{eq:resonance} and~\eqref{eq:eigen} are equivalent. However, the later form resembles the time-independent Schr{\"o}dinger equation, where the \emph{dynamical matrix} ${\mathcal D}(q)$ plays the role of a parametric Hamiltonian in the $ q$ space. 
Therefore, it is tempting to see how far this similarity extends and whether it is possible to simulate different solid-state models with interaction-coupled superlattices of BECs. 
In particular, we are interested in exploring the possibility of observing topologically nontrivial collective density excitations and flat-band-like dispersions for density oscillations. This makes Eq.~\eqref{eq:eigen} one of this letter's main \emph{formal} results and the cornerstone of what we will explore in the rest of this paper.

Considering a periodic array (i.e., superlattice) of condensates with periodic boundary condition (PBC), we can make use of the Fourier transform to find
\be\label{eq:eigen_k}
{\mathcal{D}}(q,{\bf k}) \delta \widetilde {n}(q, E,{\bf k})=E^{2}(q,{\bf k}) \delta \widetilde {n}(q, E,{\bf k}).
\ee
Here, the dynamical matrix ${\mathcal D}(q,\kv)$ has the dimension of the number of distinct condensates in each unit-cell and the dimension of ${\bf k}$-vector depends on how many spatial directions the superlattice is repeated. ${\mathcal D}$-matrix parametrically depends on the internal wave-vector $q$, as well as the superlattice wave-vector $\kv$, corresponding to the spatial periodicity of the structure. 
The internal wave vector gives the wavelength of density oscillations within each BEC site. In contrast, the superlattice wave vector $\kv$ corresponds to the phase difference between oscillations at different condensate sites. 

In principle, Eq.~\eqref{eq:eigen_k} is general and allows us to simulate different electronic Hamiltonians and, in particular, their topological properties with a suitable geometric arrangement of the BECs and the tuning of the interaction between different condensates. 
Put in another way, Eq.~\eqref{eq:eigen_k} is similar to the tight-binding Hamiltonian of a lattice model, where the on-site energy is replaced by the kinetic energy plus the intra-site interaction energy of bosons, whereas the inter-site interactions simulate the hopping of particles between different lattice sites. 

Here, we should also bear in mind a subtle difference with electronic Hamiltonians. In Eq.~\eqref{eq:eigen_k}, the excitation energy, by definition, should be non-negative, i.e., $E(q,{\bf k})\geq 0$. This enforces the ${\cal D}$-matrix to be positive semi-definitive. We will return to this point and comment on the consequences of its violation in our specific model studies.

In the following, we first showcase the most straightforward topologically nontrivial system, the SSH model with a one-dimensional BEC superlattice, and investigate its topological properties. Then, we study a two-dimensional superlattice of condensates representing a Lieb lattice supporting flat-band excitations.

\begin{figure}
	\includegraphics[width=\linewidth]{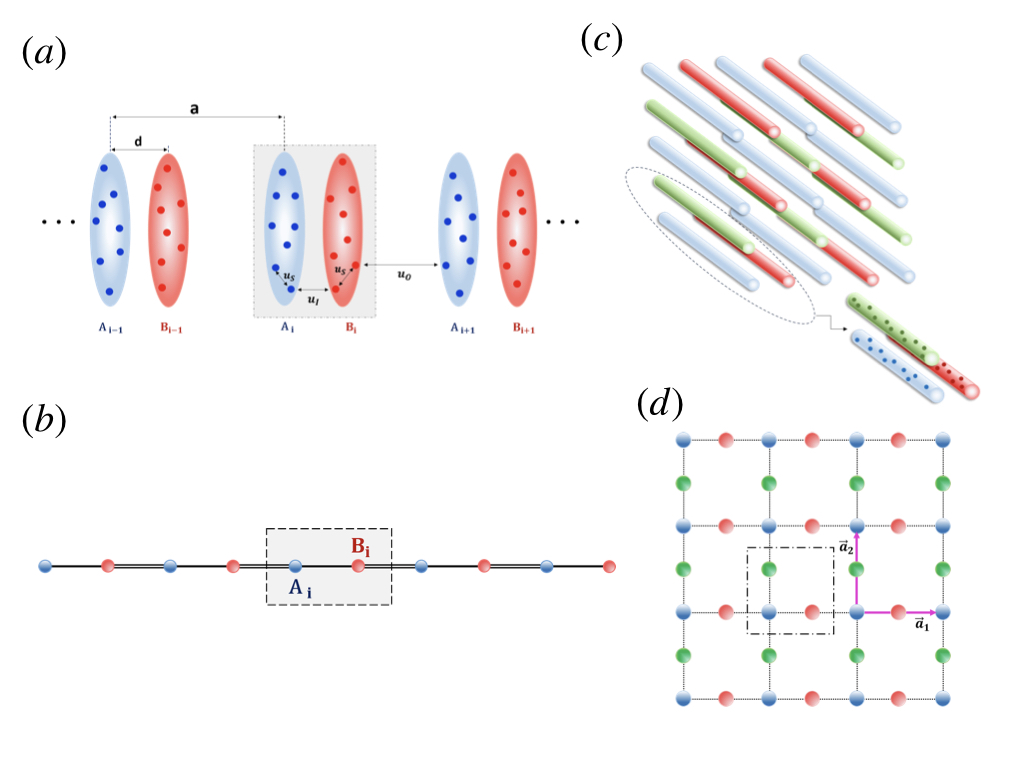}
	\caption{Schematics of different superlattices of Bose-Einstein condensates. An array of quasi-two-dimensional BECs forming a one-dimensional Su-Schrieffer-Heeger superlattice and its lattice model are shown in panels (a) and (b), respectively.
	  Panels (c) and (d) illustrate an array of quasi-one-dimensional BECs arranged into a two-dimensional Lieb superlattice and its simplified lattice model, respectively.	
	\label{fig:2D_BECs}}
\end{figure}

\emph{SSH-model of condensate superlattice.}--
Now, we consider a one-dimensional (1D) array of identical quasi-2D layers of BECs with planar densities $n$ and alternating layer spacings $d$ and $a-d$ (see, Fig.~\ref{fig:2D_BECs}, panel (a) for an illustration). With a periodic boundary condition, we have a 1D periodic structure with two layers in each unit cell of length $a$. 
Notice that as the layers are identical (i.e., with the same density and mass of particles), the $S$-matrix becomes proportional to an identity matrix, and Eq.~\eqref{eq:gen_eigen} is already a standard eigenvalue problem.

Making use of the Fourier transformation along the direction perpendicular to the 2D planes, for the elements of the $2\times 2$ dynamical matrix, we find
\be\label{d_ssh_gen}
\begin{split}
{\mathcal{D}}_{A A}(q, k)&={\mathcal{D}}_{BB}(q, k)=\varepsilon_{q}^{2}+2 n \varepsilon_{q} V_{AA}(q, k) ,\\
{\mathcal{D}}_{A B}(q, k)&={\mathcal{D}}^*_{BA}(q, k)= 2 n \varepsilon_{q}V_{AB}(q, k),
\end{split}
\ee
where $q$ is the module of the 2D in-plane wave-vector, $k$ is restricted to the 1D Brillouin zone (BZ) $[-\pi/a,\pi/a]$, and
\be
\begin{split}
V_{AA}(q, k)=& \sum_{p=-\infty}^{\infty}e^{i p ka}V_{AA}^{ii+p}(q), \\
V_{AB}(q, k)=& e^{i k d} \sum_{p=-\infty}^{+\infty} e^{i p ka} V_{AB}^{i i+p}(q) ,
\end{split}
\ee
where $V_{\alpha  \beta }^{ij}(q)$ is the interaction between particles in the $\alpha$-layer of cell $i$ and $\beta$-layer of cell $j$, and we generally have $V_{\alpha\beta}^{ii+p}(q)= V_{\beta\alpha}^{ii-p}(q)$.

In a simplified scenario, if we consider only interaction between particles in the same layer $u_{\rm S}(q)$, and nearest neighbor layers $u_{\rm I}(q)$ and $u_{\rm O}(q)$, respectively for two layers within the same cell and at two adjacent cells (see, panel (a) in Fig.~\ref{fig:2D_BECs}),   we find
\be\label{eq:d_ssh}
\begin{split}
\mathcal{D}_{A A}(q, k)&=\varepsilon_{q}^{2}+2 n \varepsilon_{q} u_{\rm S}(q), \\
\mathcal{D}_{A B}(q, k)&= 2 n \varepsilon_{q} e^{i k d}\left[u_\mathrm{I}(q)+ e^{-i k a} u_\mathrm{O}(q)\right].
\end{split}
\ee
Here, again $\mathcal{D}_{BB}=\mathcal{D}_{A A}$, and $\mathcal{D}_{ BA}=\mathcal{D}^*_{A B}$.
The eigenvalues of this ${\mathcal D}(q,k)$-matrix are the square modules of the collective mode dispersions 
\be
\begin{split}
E_{\pm}^{2}&(q, k)=\varepsilon_{q}^{2}+2 n \varepsilon_{q} u_{\mathrm{S}}(q) \\
&\pm 2 n \varepsilon_{q}\sqrt{u_{\mathrm{I}}^{2}(q)+u_{\mathrm{O}}^{2}(q)+2  u_{\mathrm{I}}(q) u_{\mathrm{O}}(q)\cos (k a) }.
\end{split}
\ee
The eigenvectors that manifest the Fourier transform of the relative amplitude of density oscillations in different layers are
\be\label{eq:bloch_SSH}
|u_\pm(q,k)\rangle=\frac{1}{\sqrt{2}}
\left(\begin{array}{c}
\pm e^{-i \phi(q,k)} \\
1
\end{array}
\right),
\ee
with
\be
\phi(q,k)\equiv \tan ^{-1}\left(\frac{u_{\mathrm{O}}(q)\sin\left(ka\right)   }{u_{\mathrm{I}}(q)+u_{\mathrm{O}}(q)\cos \left(ka\right)  }\right).
\ee

An important characteristic of collective density oscillation for bosons with short-range interactions is their sound-like long-wavelength dispersions. We can obtain the zero-sound velocities from the $q\to 0$ limits of the collective modes
\be\label{eq:zero_sound}
\begin{split}
v_{\pm}(k)&=\lim _{q \rightarrow 0} \frac{E_{\pm}(q, k)}{\hbar q} \\
&=v_s\sqrt{1\pm \sqrt{{\overline u}_{\mathrm{I}}^{2}+{\overline u}_{\mathrm{O}}^{2}+2 {\overline u}_{\mathrm{I}} {\overline u}_{\mathrm{O}} \cos (k a)}}.
\end{split}
\ee
where $v_s=\sqrt{n u_{\rm S}(q=0)/m}$ is the zero-sound velocity of an isolated layer of BEC, and ${\overline u}_{{\rm I(O)}}\equiv u_{{\rm I(O)}}(q=0)/u_{\rm S}(q=0)$ is defined for the notational brevity.
Notice that as the excitation spectrum, by definition, should always be non-negative, getting
$\varepsilon_{q}/(2 n)<\sqrt{u_{\mathrm{I}}^{2}(q)+u_{\mathrm{O}}^{2}(q)+2  u_{\mathrm{I}}(q) u_{\mathrm{O}}(q)\cos (k a) }-u_{\mathrm{S}}(q) 
$, for any values of $q$ or $k$, is a sign that the starting ground state is not the true ground state of the system. This is usually taken as a signature of the density-wave instability in the system~\cite{moudgil1997ground, Akaturk_JPC2018,Seydi_PRA2020,Pouresmaeeli_2023JPhysB}. In the following, we limit our system parameters (i.e., interaction strengths, densities, etc.) to avoid such instabilities.

\begin{figure}
	\begin{tabular}{cc}
		\includegraphics[width=0.5\linewidth]{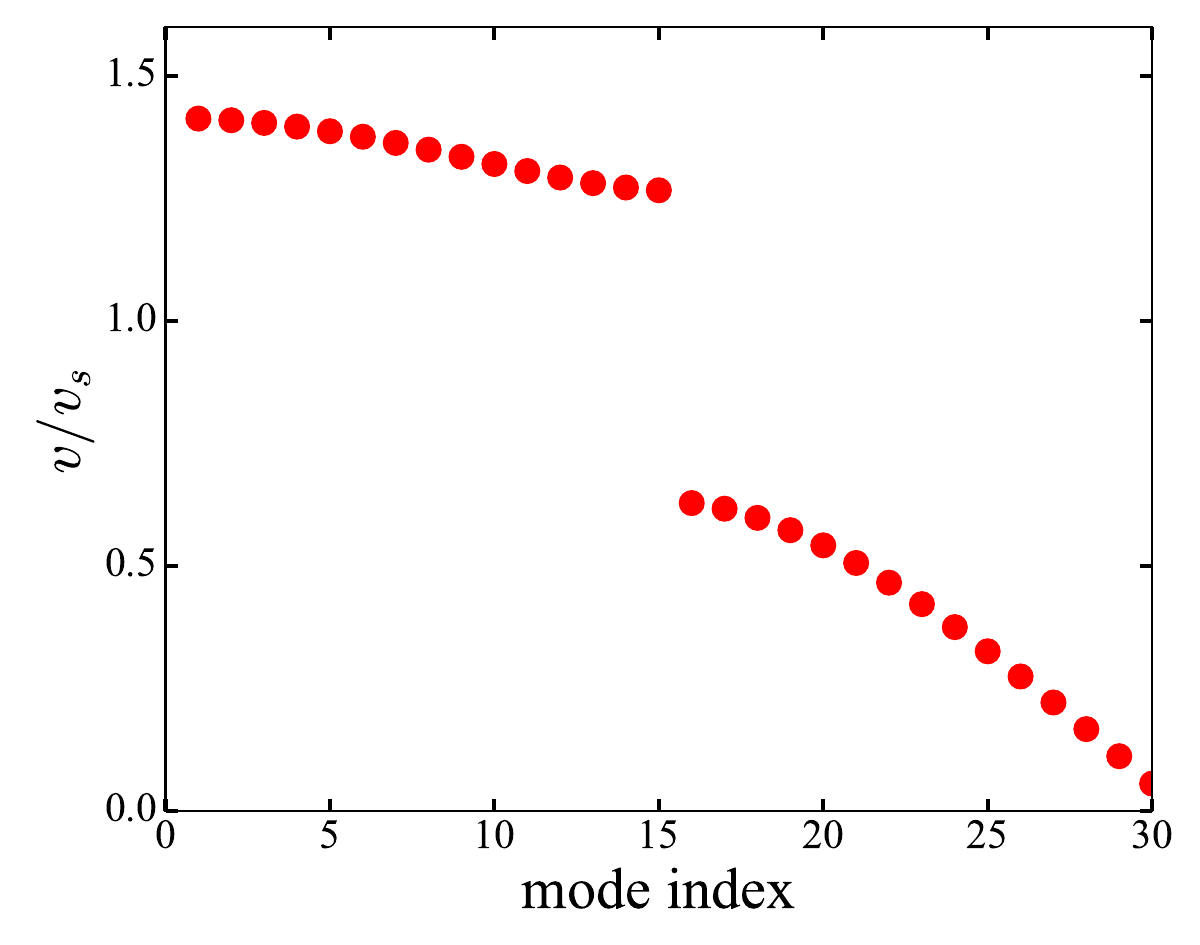} 
		\includegraphics[width=0.5\linewidth]{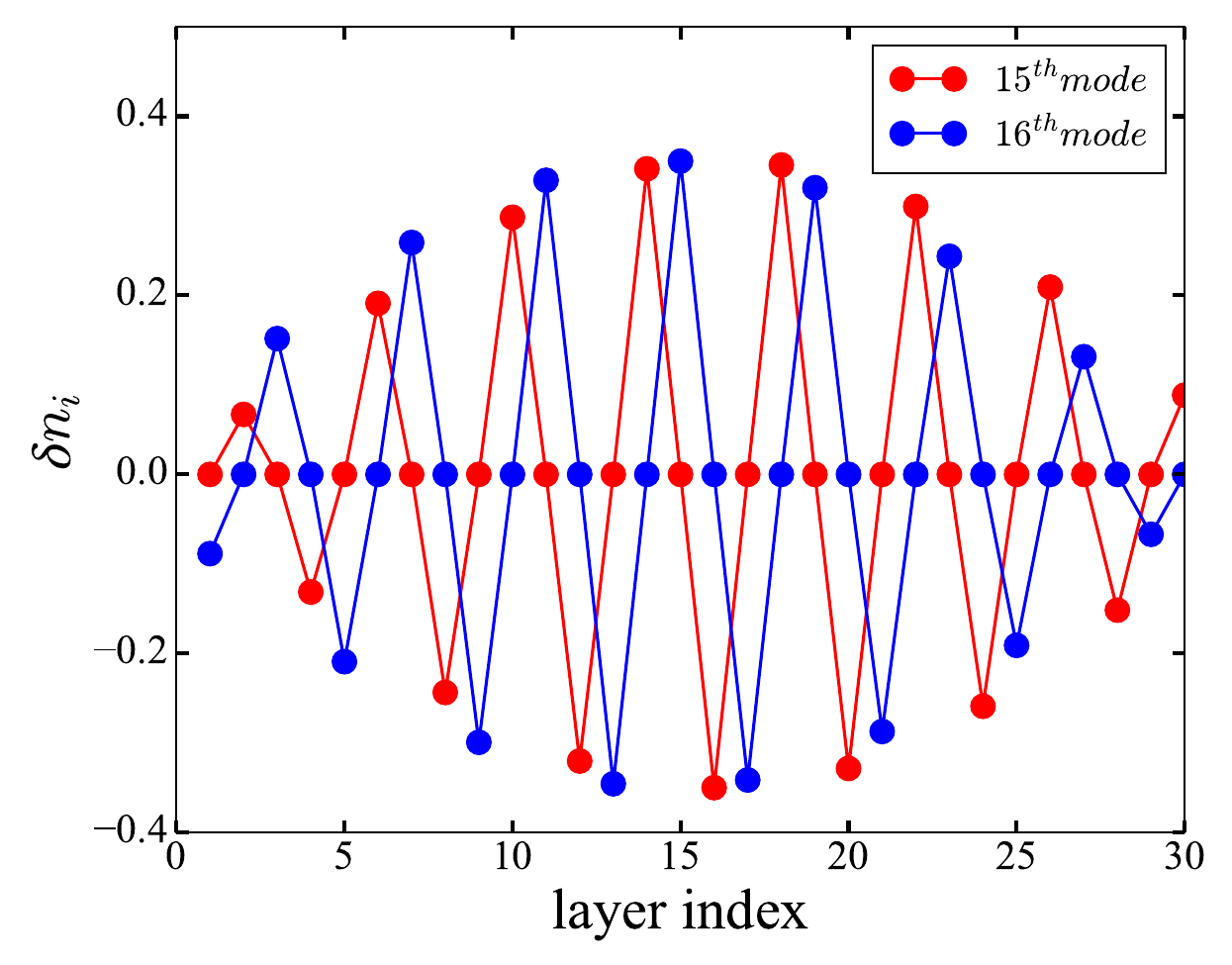} \\ 
		\includegraphics[width=0.5\linewidth]{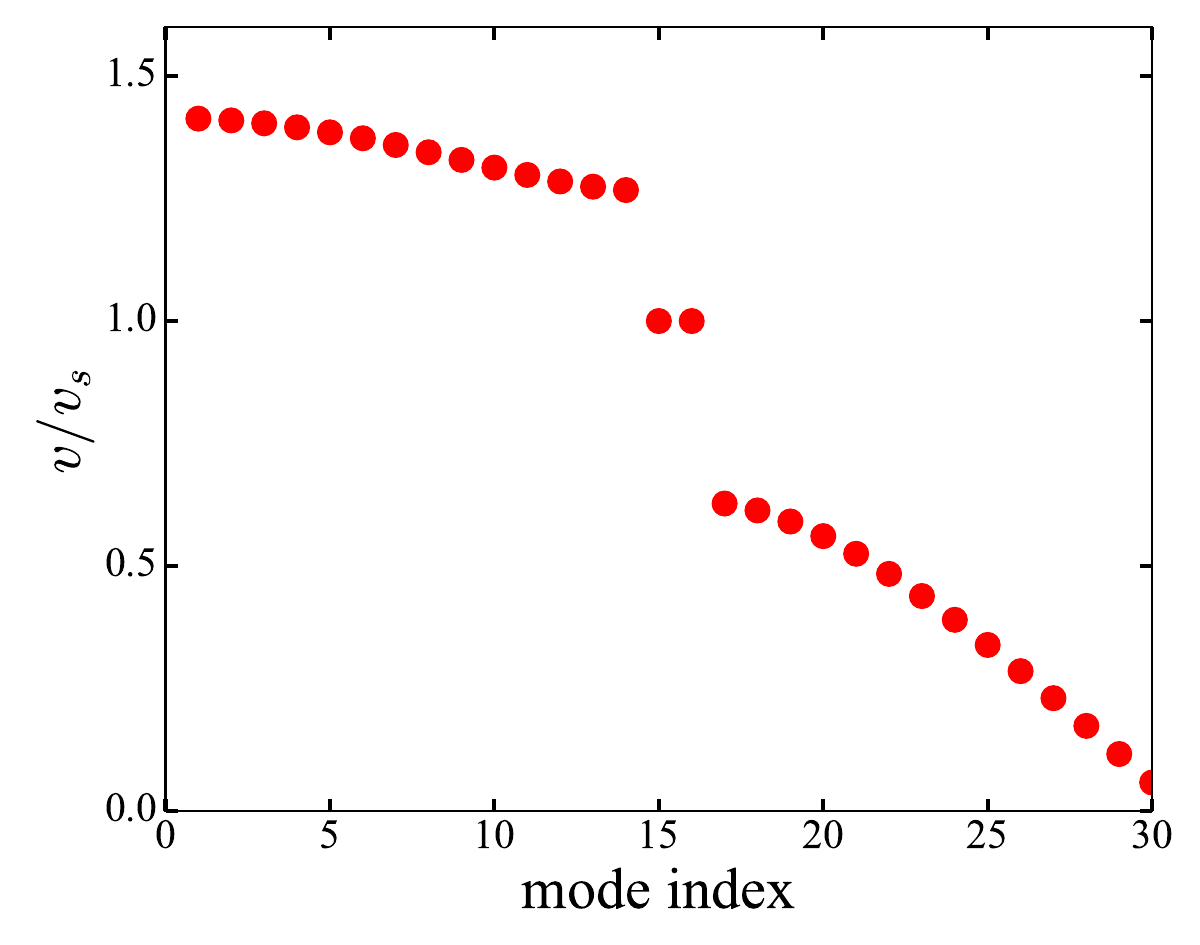}
		\includegraphics[width=0.5\linewidth]{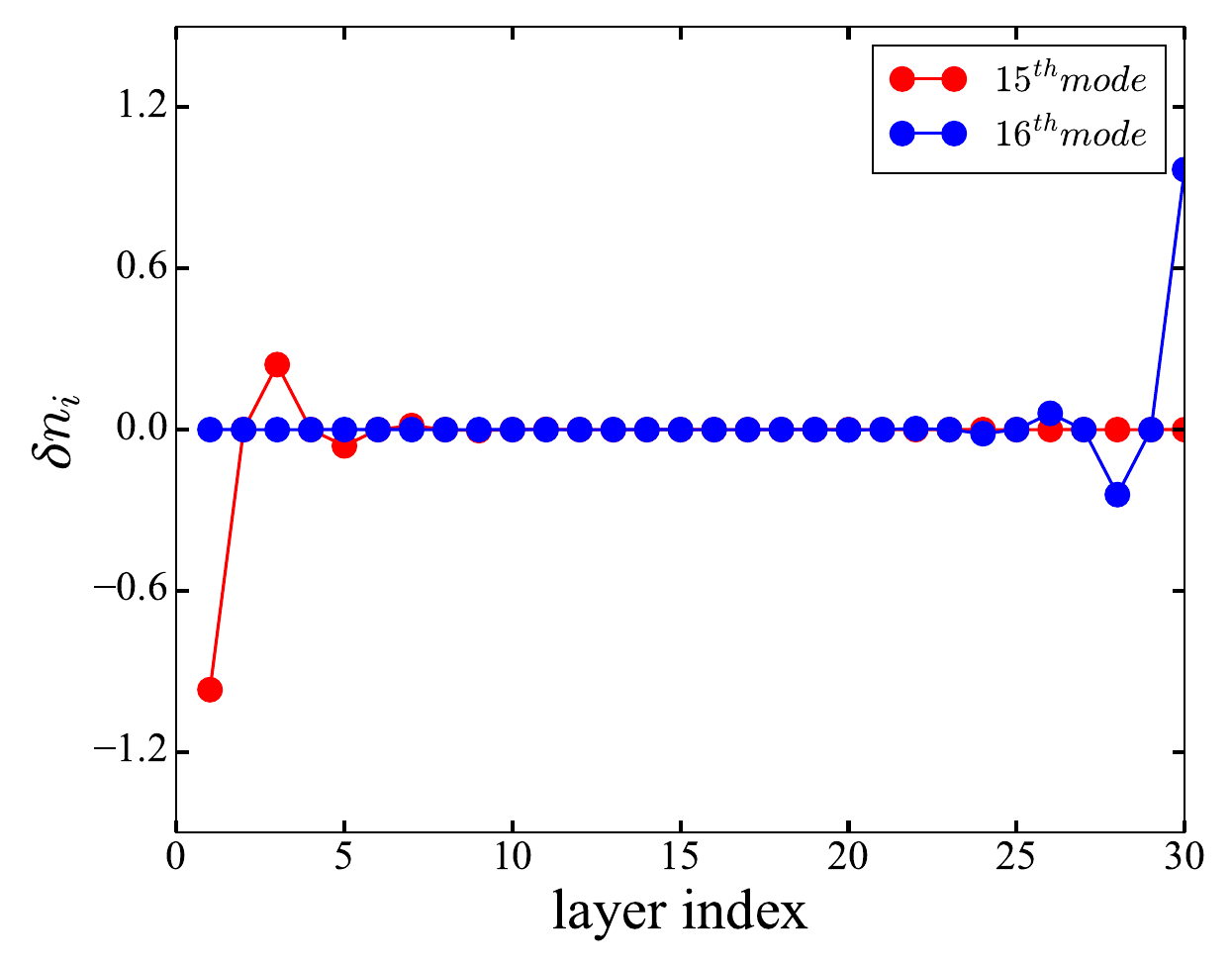}
	\end{tabular}
	\caption{Left panels: The zero-sound velocities of a finite SSH array of two-dimensional condensates, in the units of the zero-sound velocity of an isolate layer $v_s=\sqrt{n u_{\rm S}(0)/m}$. The horizontal axis shows the eigenvalue index, as sorted from the largest to the smallest value. The array consists of 15 unit cells or 30 layers.
	Right panels: The two eigenvectors of the finite SSH superlattice that correspond to the middle eigenvalues (with indices 15 and 16, for the case of 30 layers).
	The values of the interactions are $u_{\rm I}=0.8\,u_{\rm S}$, and $u_{\rm O}=0.2\, u_{\rm S}$ in the top panels and $u_{\rm I}=0.2\,u_{\rm S}$, and $u_{\rm O}=0.8\,u_{\rm S}$ in the bottom panels.
		\label{fig:zero-sound}}
\end{figure}

Note that the dynamical matrix of Eq.~\eqref{eq:d_ssh} resembles the tight-binding Hamiltonian of the one-dimensional SSH model~\cite{Asbth_Book2016}. Therefore, it is expected that they will share the same topological properties. 
The winding number of this 1D array could be defined as
\be
{\cal W}_\pm(q)=\frac{1}{2\pi }\int_{-\pi/a}^{\pi/a}{\rm d}k 
\langle u_\pm(q,k)|i\partial_k|u_\pm(q,k)\rangle,
\ee
where $u(q,k)$ are the Bloch functions of the density oscillations given by Eq.~\eqref{eq:bloch_SSH}, and $\partial_k\equiv \partial/\partial k$.
Calculating this quantity for the eigenvectors of the SSH array of condensates, we find
\be\label{eq:wq}
{\cal W}_\pm(q)=\left\{
\begin{matrix}
0 & |u_{\rm I}(q)|>|u_{\rm O}(q)|\\
1& |u_{\rm I}(q)|<|u_{\rm O}(q)|
\end{matrix}
\right. ,
\ee
and the winding number is ill-defined for $|u_{\rm I}(q)|=|u_{\rm O}(q)|$. 
In other words, the system is topologically nontrivial if the magnitude of the intra-cell $u_{\rm I}(q)$ interaction is smaller than the inter-cell interaction $u_{\rm O}(q)$. In the nontrivial state, we expect topological excitations localized at the edge layers of a finite array. 
In Fig.~\ref{fig:zero-sound} (left panels), we illustrate the results for the zero-sound velocity of a finite SSH array in the topologically trivial and nontrivial regimes. In the topological regime, we find two modes whose sound velocities are identical to the sound velocity of an isolated layer. 
The amplitude of the density oscillations corresponding to the middle eigenvalues reveals that in the topologically state, the density oscillations are localized in the edge layers. However, they are extended throughout the chain in a trivial system (Fig.~\ref{fig:zero-sound}, right panels).  

\begin{figure}
	\begin{tabular}{cc}
	\includegraphics[width=0.5\linewidth]{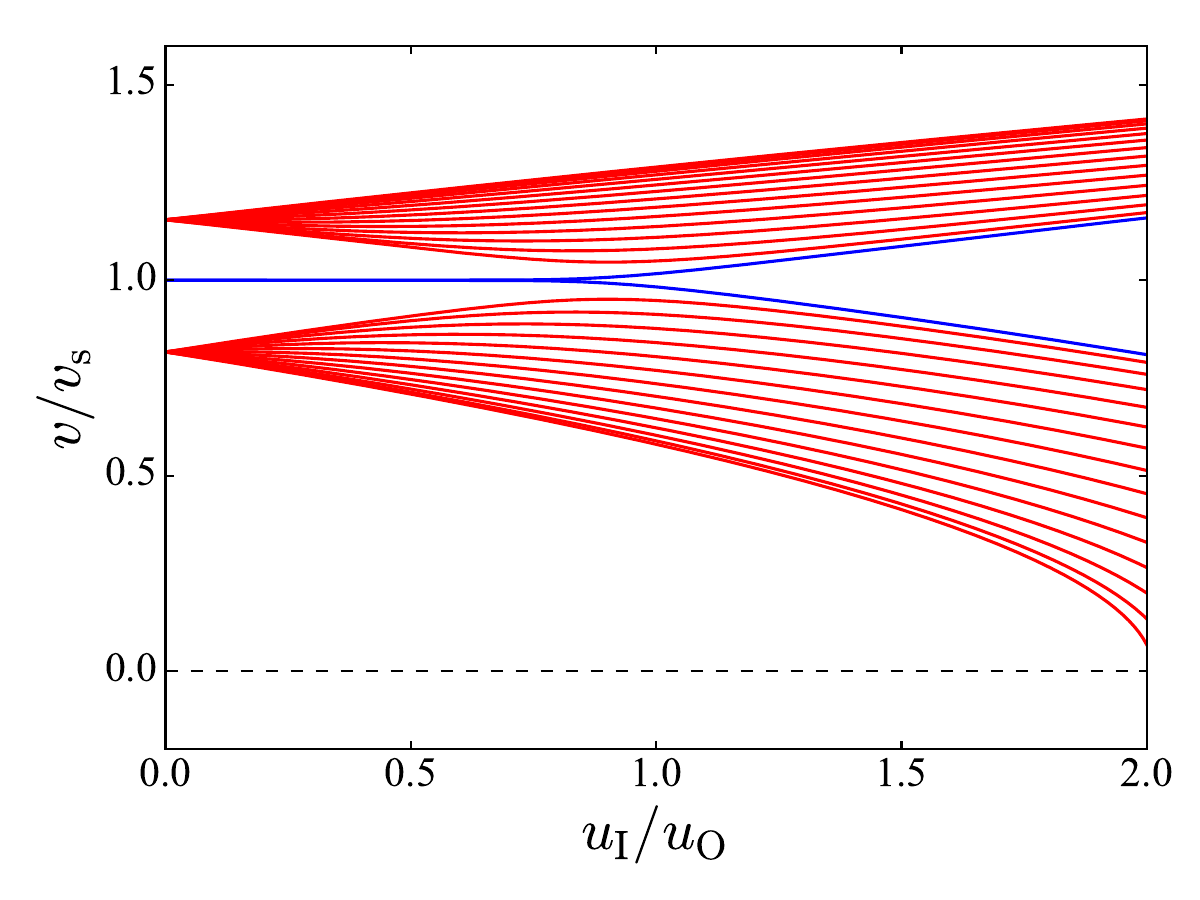}  
	\includegraphics[width=0.5\linewidth]{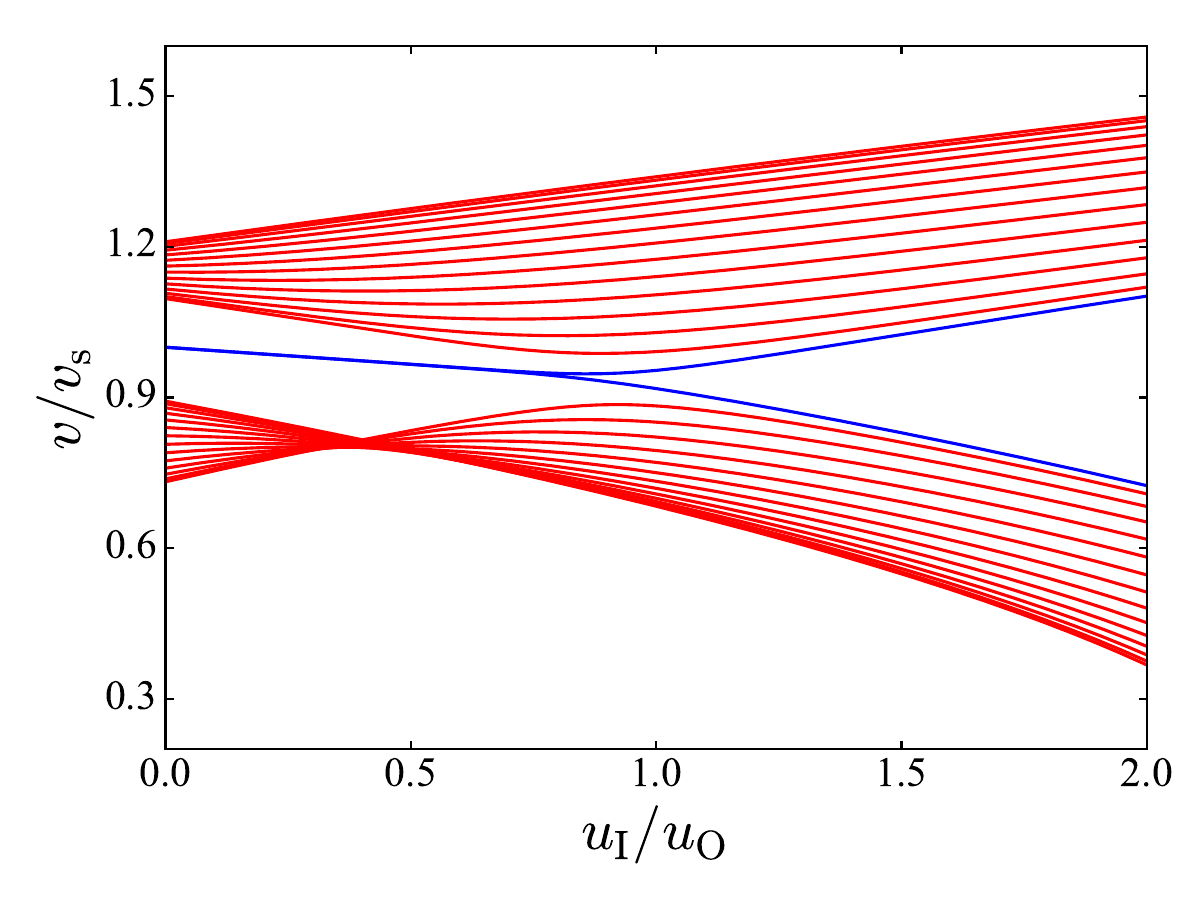}
	\end{tabular}
	\caption{Left: The zero-sound velocities of a finite SSH array of 2D condensates with 15 unit cells versus the ratio of the interaction between nearest neighbor layers. The blue curves identify the middle (in this case, corresponding to the $15^{\rm th}$ and $16^{\rm st}$) eigenvalues. 
	Right: Same as the left panel, but the effects of the interaction between the second neighbor layers (with $u_n=0.2\,u_{\rm S}$) are also included. \label{fig:zero-sound2}}
\end{figure}

In Fig.~\ref{fig:zero-sound2}, we show the transition from the topologically nontrivial to the trivial phase as the ratio between intra-cell and inter-cell interactions varies. The effect of interaction between further away neighbors is also illustrated in the right panel of the same figure~\cite{supplemental}.

\emph{Wave-vector dependent winding number.}--
Interestingly, the winding number defined through Eq.~\eqref{eq:wq} depends on the in-plane wave vector $q$. To better explore the consequences of this $q$-dependence, we consider an SSH array of condensates with a soft-core interaction between particles $u(r_\parallel)=U\Theta ({r_c} - r)$, where $U$ is the interaction strength, $\Theta(x)$ is the Heaviside step function, and $r_c$ is the soft-core radius of the interaction. The distance between two particles is $r$, while $r_\parallel$ refers to their in-plane (2D) separation.
If we choose the system parameters such that ${\rm max}\{d, a-d\}<r_c<a$, then just intra-layer and nearest neighbor inter-layer interactions are non-zero, and the system exactly corresponds to an SSH array with nearest neighbor couplings.
After performing the 2D Fourier transforms of the soft-core step interaction, we find
$u_{\rm S}(q) = 2\pi U r_c J_1(r_c q)/q$
where ${J_1}(x)$ is the Bessel function of the first kind~\cite{Abramowitz_Book1965}. We can find $u_{\rm I}(q)$ and $u_{\rm O}(q)$ from $u_{\rm S}(q)$  by replacing $r_c$ with $\sqrt {r_c^2 - d^2} $ and $\sqrt {r_c^2 - (a-d)^2} $, respectively. 

\begin{figure}
	\begin{tabular}{cc}
	\includegraphics[width=0.5\linewidth]{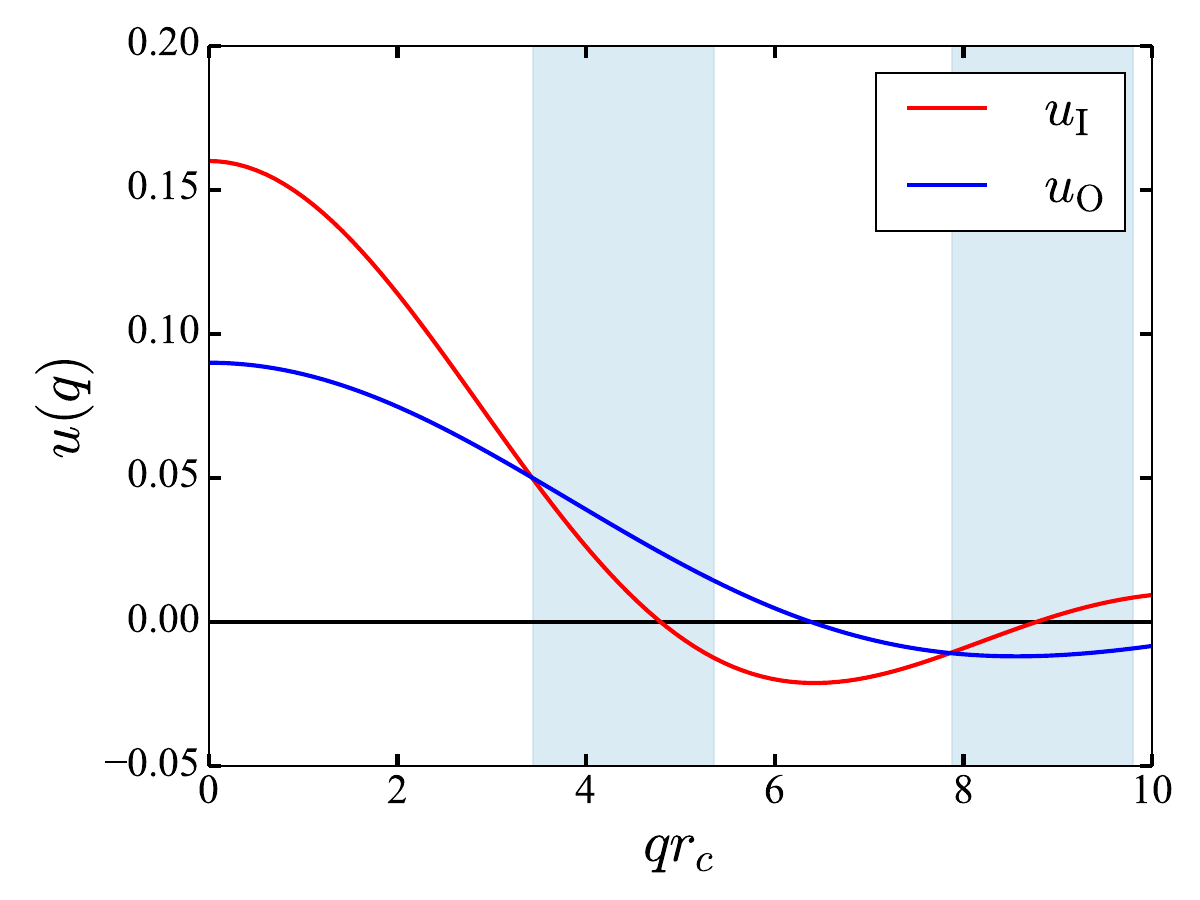}
	\includegraphics[width=0.5\linewidth]{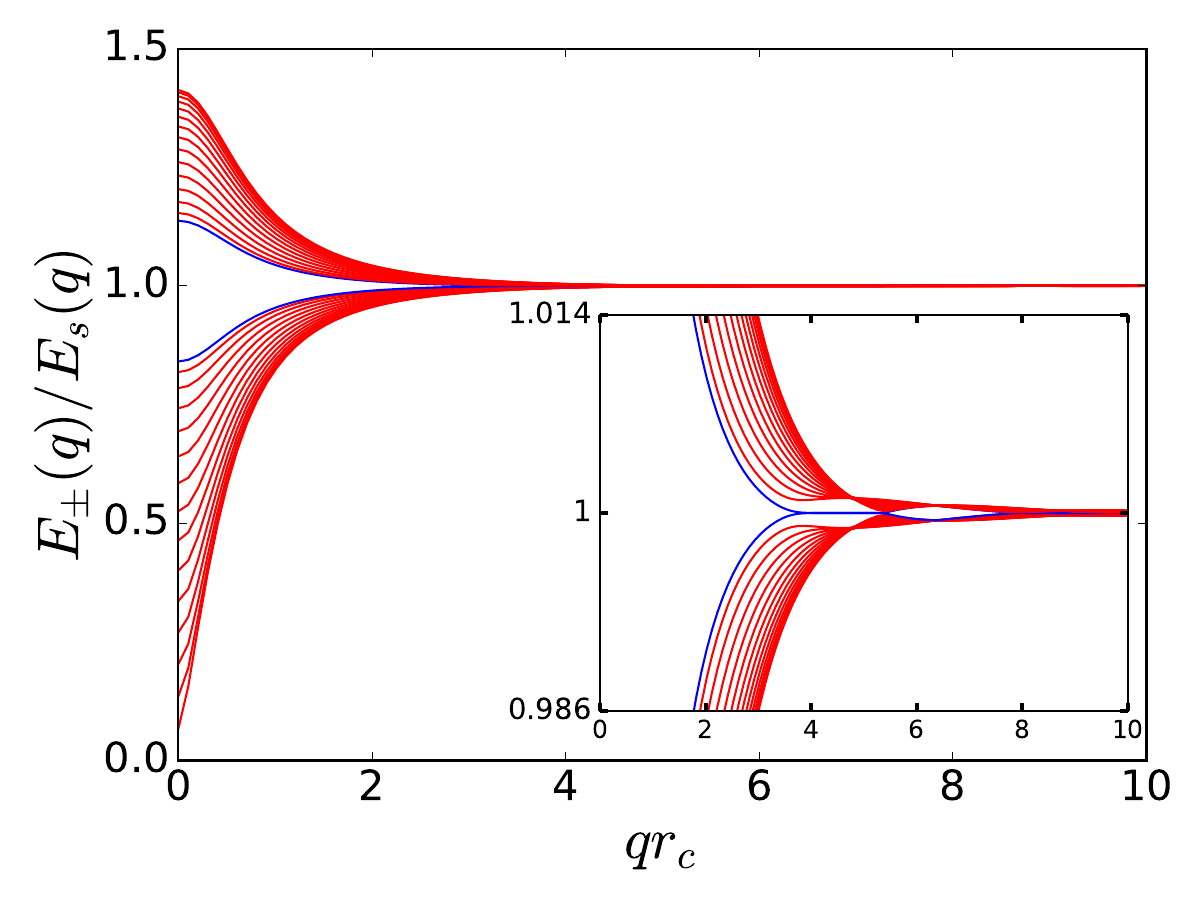}
	\end{tabular}
	\caption{Left panel: The Fourier transforms of the intra-cell and inter-cell short-range soft-core interactions [in the units of $m/(2\pi\hbar^{2})$]. In the shaded regions, the winding number is non-zero.
	Right panel: The dispersion of the excitation spectrum of a finite SSH superlattice of condensates with short-range soft-core interaction and 15 unit cells, plotted in the units of the excitation spectrum of an isolated layer $E_s( q)=\sqrt{\varepsilon_{ q}^{2}+2 n \varepsilon_{ q} u_{\rm S}(q)}$.  
	The blue curves highlight the dispersions of the middle eigenvalues.
	The inset in the right panel shows the zoom into the energies close to $E_s(q)$ to better illustrate the topological edge modes.
	The unit cell constant and layer spacing values are $a=1.4\, r_c$ and $d=0.6\, r_c $, respectively. Also, the value of interaction strength is considered to be $U=2 m r_c^2/\hbar^2$.
		 	\label{fig:soft}}
\end{figure}

In Fig.~\ref{fig:soft} (left panel), we have plotted the typical wave vector dependence of the intra-cell and inter-cell components of the soft-core interaction. The excitation spectrum $E(q)$ versus the wave vector of a finite sub-lattice under such interactions is illustrated in the right panel of Fig.~\ref{fig:soft}.  
We observe that the winding number is non-zero just for some ranges of the wave vectors (the shaded area in the right panel). In other words, the system supports localized edge excitations for some values of the wave vectors, but all the excitations are extended across the array for other wave vectors. 
This is an interesting phenomenon with no similar counterpart in solid-state electronic topological systems. 
We expect that such localized collective modes would be vulnerable against inelastic scatterings.

\emph{Two-dimensional Lieb superlattice and flat-band excitations.---}
Electronic flat bands have recently attracted considerable interest in condensed matter physics, as they are responsible for several interesting emergent phenomena~\cite{Regnault2022,Leykam2018,chen_arxiv2024}. A prototypical tight-binding model that supports such an electronic flat band is a two-dimensional Lieb lattice.  

In the following, we consider an array of quasi-one-dimensional BEC tubes arranged in a two-dimensional Lieb lattice structure (see right panels in Fig.~\ref{fig:2D_BECs}). 
Making use of the Fourier transformation along the two directions perpendicular to the tube axis for the elements of
the $3 \times 3$ ${\cal D}$-matrix we find
\be
 {\cal D}_{\alpha\beta}(q,\kv)=\varepsilon^2_{q,\alpha}  \delta_{\alpha,\beta}
+2 \sqrt{n_{\alpha}  n_{\beta }\varepsilon_{q, \alpha} \varepsilon_{q, \beta}}\, V_{\alpha\beta}(q,\kv),
\ee
where $\alpha,\beta=A,B$, and $C$, refer to three different sub-lattices in the Lieb lattice.
In the simplified model, considering identical tubes of condensate and short-range interactions, we find 
\be\label{eq:D_Lieb}
\begin{split}
&	{ {\cal D}}_{\alpha\alpha}=\varepsilon^2_{q}  +2 n \varepsilon_{q} u_{\rm S}(q) ,\\
&	{ {\cal D}}_{AB}={ {\cal D}}_{BA}=4 n \varepsilon_{q} u_{1}(q) \cos(k_xa/2),\\
&	{ {\cal D}}_{AC}={ {\cal D}}_{CA}=4 n \varepsilon_{q} u_{1}(q)\cos(k_xa/2),\\
&	{ {\cal D}}_{BC}={ {\cal D}}_{CB}=8 n \varepsilon_{q} u_{2}(q) \cos(k_xa/2)\cos(k_ya/2).
\end{split}
\ee
Here, $u_{\rm S}(q)$ is the Fourier transform of the intra-tube interaction, and $u_{1}(q)$ and $u_{2}(q)$ refer to the Fourier transforms of the nearest neighbor and next-nearest neighbor inter-tube interactions, respectively. 
The interactions between more distant neighbors are ignored for brevity. 

If we further discard the next-nearest neighbor interactions (i.e., $u_2(q)\approx 0$), we analytically find the eigenvalues of the dynamical matrix for Lieb lattice as
\be
\begin{split}
E^2_0(q,\kv)&=\varepsilon^2_{q}+2 n\varepsilon_{q} u_{\rm S}(q), \\
E^2_\pm (q,\kv) &=\varepsilon^2_{q}+2 n\varepsilon_{q} u_{\rm S}(q)  \pm 4 n \varepsilon_{q}  u_1(q) \left| {\bf d}(\kv)\right|,
\end{split}
\ee
where ${\bf d}(\kv)=\left(\cos({k_{x}a}/2),\cos({k_{y}a}/{2})\right)$, is a 2D vector defined for the notational brevity.
The first eigenvalue $E_0$ is identical to the Bogoliubov dispersion of an isolated BEC. Its superlattice wave-vector $\kv$ independence resembles the electronic flat-bands in solid-state models. 
In other words, there is a collective density mode unaffected by the inter-tube interactions for any value of the inner wave vector $q$.

\begin{figure}
	\begin{tabular}{cc}
	\includegraphics[width=0.5\linewidth]{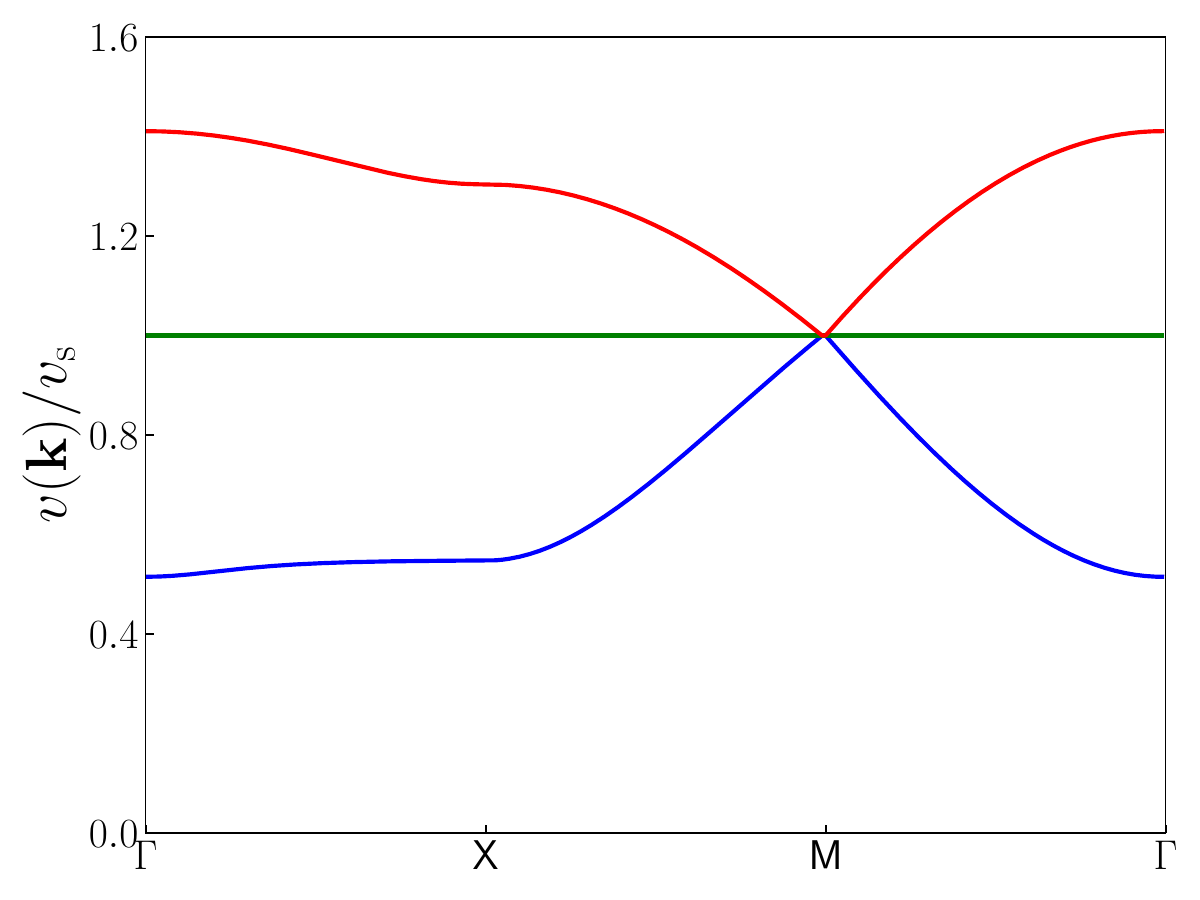}
	 \includegraphics[width=0.5\linewidth]{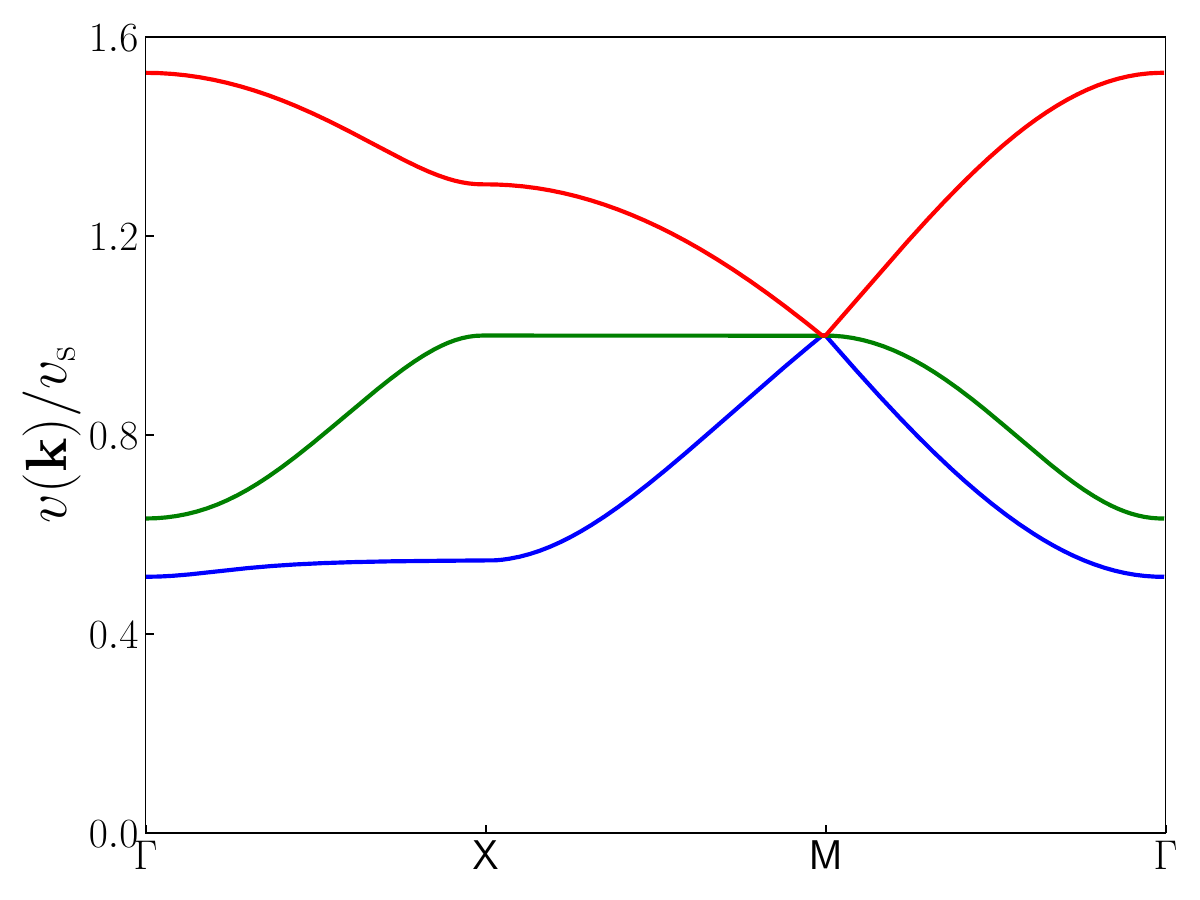}
	\end{tabular}
	\caption{The dispersions of the zero-sound velocities (in the units of $v_s$) versus $\kv$, along the high-symmetry path $\Gamma-X-M-\Gamma$ of the two-dimensional BZ. 
	The long wavelength magnitude of the nearest neighbor interaction is fixed at $u_1(0)=0.35\, u_{\rm S}(0)$. The next-nearest-neighbor interaction is $u_2(0)=0$ (left panel)  and $u_2(0)=0.15\, u_{\rm S}(0)$ (right panel).
		\label{fig:3D_Lieb}}
\end{figure}

In the long wavelengths limit, we find three different zero-sound velocities, one with $v_0=v_s=\sqrt{nu_{\rm S}(0)/m}$, that is same as the sound velocity of an isolated 1D BEC tube, and two other modes with $v_{\pm}(\kv)=v_s\sqrt{1 \pm 2  u_1(0) \left| {\bf d}(\kv)\right|/u_{\rm S}(0)}$. 
In Fig.~\ref{fig:3D_Lieb}, we show three zero-sound velocities versus the superlattice wave vector $\kv$ in a 2D Lieb lattice, where the $\kv$-independent sound mode is evident. The effects of next-nearest-neighbor interaction $u_2$ are also shown in the right panel of Fig.~\ref{fig:3D_Lieb}. Note that for finite $u_2$, we obtain the eigenvalues from the numerical diagonalization of the ${\cal D}$-matrix in Eq.~\eqref{eq:D_Lieb}.
All three velocities are degenerate at the corners of the first BZ, i.e., $M=(\pi/a,\pi/a)$. Furthermore, along the edges of the BZ, where either $k_x$ or $k_y$ is $\pm \pi/a$, the middle mode maintains its flat-band nature even in the presence of the next nearest neighbor interaction (see the right panel in Fig.~\ref{fig:3D_Lieb}).

\emph{Conclusions and future directions.}--
Starting from the linear response equation for the density oscillations of an array of Bose-Einstein condensates, we showed that an eigenvalue equation gives the dispersions of the system's collective density oscillations.
This representation in the eigenvalue problem form enables us to explore the similarities between different tight-binding lattice models and superlattices of interaction-coupled Bose-Einstein condensates in an obvious way. 
In particular, in this letter, we investigated the possibility of observing topologically nontrivial edge excitations and flat-band modes with a suitably engineered superlattice of BECs. 
This proposal provides a new avenue for simulating quantum materials with ultra-cold atoms in optical lattices~\cite{Schfer_NRP2020}.  

As an illustrative example for our formulation, we showed how the Su-Schrieffer-Heeger model of the condensates could be realized. Then we obtained its edge excitations in the topologically nontrivial regime. 
It seems straightforward to investigate other more exotic topological models as well. 
For two-dimensional models, it is possible to think of an array of quasi-one-dimensional condensates. As a concrete example, we demonstrated how flat-band-like density modes could be realized in a two-dimensional Lieb superlattice structure of BECs.

The effects of finite temperature and dissipation on our findings need further investigation. Including the quantum fluctuations beyond our mean-field treatment also remains to be explored. 
Furthermore, it would be interesting to see how similar ideas could be extended to fermions, where superlattice structures could be naturally realized in layered materials.

\emph{Acknowledgments.}--
We thank B. Tanatar, Ali G. Moghaddam, and Jahanfar Abouie for valuable discussions.
This project is supported by the Iran Science Elites Federation (ISEF) and the Research Council of the Institute for Advanced Studies in Basic Sciences (IASBS).

\bibliography{topo_BEC.bbl}

\begin{thebibliography}{29}%
\makeatletter
\providecommand \@ifxundefined [1]{%
 \@ifx{#1\undefined}
}%
\providecommand \@ifnum [1]{%
 \ifnum #1\expandafter \@firstoftwo
 \else \expandafter \@secondoftwo
 \fi
}%
\providecommand \@ifx [1]{%
 \ifx #1\expandafter \@firstoftwo
 \else \expandafter \@secondoftwo
 \fi
}%
\providecommand \natexlab [1]{#1}%
\providecommand \enquote  [1]{``#1''}%
\providecommand \bibnamefont  [1]{#1}%
\providecommand \bibfnamefont [1]{#1}%
\providecommand \citenamefont [1]{#1}%
\providecommand \href@noop [0]{\@secondoftwo}%
\providecommand \href [0]{\begingroup \@sanitize@url \@href}%
\providecommand \@href[1]{\@@startlink{#1}\@@href}%
\providecommand \@@href[1]{\endgroup#1\@@endlink}%
\providecommand \@sanitize@url [0]{\catcode `\\12\catcode `\$12\catcode
  `\&12\catcode `\#12\catcode `\^12\catcode `\_12\catcode `\%12\relax}%
\providecommand \@@startlink[1]{}%
\providecommand \@@endlink[0]{}%
\providecommand \url  [0]{\begingroup\@sanitize@url \@url }%
\providecommand \@url [1]{\endgroup\@href {#1}{\urlprefix }}%
\providecommand \urlprefix  [0]{URL }%
\providecommand \Eprint [0]{\href }%
\providecommand \doibase [0]{https://doi.org/}%
\providecommand \selectlanguage [0]{\@gobble}%
\providecommand \bibinfo  [0]{\@secondoftwo}%
\providecommand \bibfield  [0]{\@secondoftwo}%
\providecommand \translation [1]{[#1]}%
\providecommand \BibitemOpen [0]{}%
\providecommand \bibitemStop [0]{}%
\providecommand \bibitemNoStop [0]{.\EOS\space}%
\providecommand \EOS [0]{\spacefactor3000\relax}%
\providecommand \BibitemShut  [1]{\csname bibitem#1\endcsname}%
\let\auto@bib@innerbib\@empty
\bibitem [{\citenamefont {Asbóth}\ \emph {et~al.}(2016)\citenamefont
  {Asbóth}, \citenamefont {Oroszlány},\ and\ \citenamefont
  {Pályi}}]{Asbth_Book2016}%
  \BibitemOpen
  \bibfield  {author} {\bibinfo {author} {\bibfnamefont {J.~K.}\ \bibnamefont
  {Asbóth}}, \bibinfo {author} {\bibfnamefont {L.}~\bibnamefont
  {Oroszlány}},\ and\ \bibinfo {author} {\bibfnamefont {A.}~\bibnamefont
  {Pályi}},\ }\href {https://doi.org/10.1007/978-3-319-25607-8} {\emph
  {\bibinfo {title} {A Short Course on Topological Insulators}}}\ (\bibinfo
  {publisher} {Springer International Publishing},\ \bibinfo {year}
  {2016})\BibitemShut {NoStop}%
\bibitem [{\citenamefont {Moessner}\ and\ \citenamefont
  {Moore}(2021)}]{Moessner2021}%
  \BibitemOpen
  \bibfield  {author} {\bibinfo {author} {\bibfnamefont {R.}~\bibnamefont
  {Moessner}}\ and\ \bibinfo {author} {\bibfnamefont {J.~E.}\ \bibnamefont
  {Moore}},\ }\href {https://doi.org/10.1017/9781316226308} {\emph {\bibinfo
  {title} {Topological Phases of Matter}}}\ (\bibinfo  {publisher} {Cambridge
  University Press},\ \bibinfo {year} {2021})\BibitemShut {NoStop}%
\bibitem [{\citenamefont {Regnault}\ \emph {et~al.}(2022)\citenamefont
  {Regnault}, \citenamefont {Xu}, \citenamefont {Li}, \citenamefont {Ma},
  \citenamefont {Jovanovic}, \citenamefont {Yazdani}, \citenamefont {Parkin},
  \citenamefont {Felser}, \citenamefont {Schoop}, \citenamefont {Ong},
  \citenamefont {Cava}, \citenamefont {Elcoro}, \citenamefont {Song},\ and\
  \citenamefont {Bernevig}}]{Regnault2022}%
  \BibitemOpen
  \bibfield  {author} {\bibinfo {author} {\bibfnamefont {N.}~\bibnamefont
  {Regnault}}, \bibinfo {author} {\bibfnamefont {Y.}~\bibnamefont {Xu}},
  \bibinfo {author} {\bibfnamefont {M.-R.}\ \bibnamefont {Li}}, \bibinfo
  {author} {\bibfnamefont {D.-S.}\ \bibnamefont {Ma}}, \bibinfo {author}
  {\bibfnamefont {M.}~\bibnamefont {Jovanovic}}, \bibinfo {author}
  {\bibfnamefont {A.}~\bibnamefont {Yazdani}}, \bibinfo {author} {\bibfnamefont
  {S.~S.~P.}\ \bibnamefont {Parkin}}, \bibinfo {author} {\bibfnamefont
  {C.}~\bibnamefont {Felser}}, \bibinfo {author} {\bibfnamefont {L.~M.}\
  \bibnamefont {Schoop}}, \bibinfo {author} {\bibfnamefont {N.~P.}\
  \bibnamefont {Ong}}, \bibinfo {author} {\bibfnamefont {R.~J.}\ \bibnamefont
  {Cava}}, \bibinfo {author} {\bibfnamefont {L.}~\bibnamefont {Elcoro}},
  \bibinfo {author} {\bibfnamefont {Z.-D.}\ \bibnamefont {Song}},\ and\
  \bibinfo {author} {\bibfnamefont {B.~A.}\ \bibnamefont {Bernevig}},\
  }\bibfield  {title} {\bibinfo {title} {Catalogue of flat-band stoichiometric
  materials},\ }\href {https://doi.org/10.1038/s41586-022-04519-1} {\bibfield
  {journal} {\bibinfo  {journal} {Nature}\ }\textbf {\bibinfo {volume} {603}},\
  \bibinfo {pages} {824–828} (\bibinfo {year} {2022})}\BibitemShut {NoStop}%
\bibitem [{\citenamefont {Krishnamoorthy}\ \emph {et~al.}(2023)\citenamefont
  {Krishnamoorthy}, \citenamefont {Dubrovkin}, \citenamefont {Adamo},\ and\
  \citenamefont {Soci}}]{Krishnamoorthy2023}%
  \BibitemOpen
  \bibfield  {author} {\bibinfo {author} {\bibfnamefont {H.~N.~S.}\
  \bibnamefont {Krishnamoorthy}}, \bibinfo {author} {\bibfnamefont {A.~M.}\
  \bibnamefont {Dubrovkin}}, \bibinfo {author} {\bibfnamefont {G.}~\bibnamefont
  {Adamo}},\ and\ \bibinfo {author} {\bibfnamefont {C.}~\bibnamefont {Soci}},\
  }\bibfield  {title} {\bibinfo {title} {Topological insulator metamaterials},\
  }\href {https://doi.org/10.1021/acs.chemrev.2c00594} {\bibfield  {journal}
  {\bibinfo  {journal} {Chemical Reviews}\ }\textbf {\bibinfo {volume} {123}},\
  \bibinfo {pages} {4416–4442} (\bibinfo {year} {2023})}\BibitemShut
  {NoStop}%
\bibitem [{\citenamefont {Leykam}\ \emph {et~al.}(2018)\citenamefont {Leykam},
  \citenamefont {Andreanov},\ and\ \citenamefont {Flach}}]{Leykam2018}%
  \BibitemOpen
  \bibfield  {author} {\bibinfo {author} {\bibfnamefont {D.}~\bibnamefont
  {Leykam}}, \bibinfo {author} {\bibfnamefont {A.}~\bibnamefont {Andreanov}},\
  and\ \bibinfo {author} {\bibfnamefont {S.}~\bibnamefont {Flach}},\ }\bibfield
   {title} {\bibinfo {title} {Artificial flat band systems: from lattice models
  to experiments},\ }\href {https://doi.org/10.1080/23746149.2018.1473052}
  {\bibfield  {journal} {\bibinfo  {journal} {Advances in Physics: X}\ }\textbf
  {\bibinfo {volume} {3}},\ \bibinfo {pages} {1473052} (\bibinfo {year}
  {2018})}\BibitemShut {NoStop}%
\bibitem [{\citenamefont {Ma}\ \emph {et~al.}(2019)\citenamefont {Ma},
  \citenamefont {Xiao},\ and\ \citenamefont {Chan}}]{ma2019topological}%
  \BibitemOpen
  \bibfield  {author} {\bibinfo {author} {\bibfnamefont {G.}~\bibnamefont
  {Ma}}, \bibinfo {author} {\bibfnamefont {M.}~\bibnamefont {Xiao}},\ and\
  \bibinfo {author} {\bibfnamefont {C.~T.}\ \bibnamefont {Chan}},\ }\bibfield
  {title} {\bibinfo {title} {Topological phases in acoustic and mechanical
  systems},\ }\href {https://doi.org/10.1038/s42254-019-0030-x} {\bibfield
  {journal} {\bibinfo  {journal} {Nature Reviews Physics}\ }\textbf {\bibinfo
  {volume} {1}},\ \bibinfo {pages} {281} (\bibinfo {year} {2019})}\BibitemShut
  {NoStop}%
\bibitem [{\citenamefont {Many~Manda}\ \emph {et~al.}(2022)\citenamefont
  {Many~Manda}, \citenamefont {Chaunsali}, \citenamefont {Theocharis},\ and\
  \citenamefont {Skokos}}]{manda2022nonlinear}%
  \BibitemOpen
  \bibfield  {author} {\bibinfo {author} {\bibfnamefont {B.}~\bibnamefont
  {Many~Manda}}, \bibinfo {author} {\bibfnamefont {R.}~\bibnamefont
  {Chaunsali}}, \bibinfo {author} {\bibfnamefont {G.}~\bibnamefont
  {Theocharis}},\ and\ \bibinfo {author} {\bibfnamefont {C.}~\bibnamefont
  {Skokos}},\ }\bibfield  {title} {\bibinfo {title} {Nonlinear topological edge
  states: From dynamic delocalization to thermalization},\ }\href
  {https://doi.org/10.1103/PhysRevB.105.104308} {\bibfield  {journal} {\bibinfo
   {journal} {Phys. Rev. B}\ }\textbf {\bibinfo {volume} {105}},\ \bibinfo
  {pages} {104308} (\bibinfo {year} {2022})}\BibitemShut {NoStop}%
\bibitem [{\citenamefont {Haldane}\ and\ \citenamefont
  {Raghu}(2008)}]{Haldane_PRL2008}%
  \BibitemOpen
  \bibfield  {author} {\bibinfo {author} {\bibfnamefont {F.~D.~M.}\
  \bibnamefont {Haldane}}\ and\ \bibinfo {author} {\bibfnamefont
  {S.}~\bibnamefont {Raghu}},\ }\bibfield  {title} {\bibinfo {title} {Possible
  realization of directional optical waveguides in photonic crystals with
  broken time-reversal symmetry},\ }\href
  {https://doi.org/10.1103/PhysRevLett.100.013904} {\bibfield  {journal}
  {\bibinfo  {journal} {Phys. Rev. Lett.}\ }\textbf {\bibinfo {volume} {100}},\
  \bibinfo {pages} {013904} (\bibinfo {year} {2008})}\BibitemShut {NoStop}%
\bibitem [{\citenamefont {Raghu}\ and\ \citenamefont
  {Haldane}(2008)}]{Raghu_PRA2008}%
  \BibitemOpen
  \bibfield  {author} {\bibinfo {author} {\bibfnamefont {S.}~\bibnamefont
  {Raghu}}\ and\ \bibinfo {author} {\bibfnamefont {F.~D.~M.}\ \bibnamefont
  {Haldane}},\ }\bibfield  {title} {\bibinfo {title} {Analogs of
  quantum-hall-effect edge states in photonic crystals},\ }\href
  {https://doi.org/10.1103/PhysRevA.78.033834} {\bibfield  {journal} {\bibinfo
  {journal} {Phys. Rev. A}\ }\textbf {\bibinfo {volume} {78}},\ \bibinfo
  {pages} {033834} (\bibinfo {year} {2008})}\BibitemShut {NoStop}%
\bibitem [{\citenamefont {Jiang}\ \emph {et~al.}(2021)\citenamefont {Jiang},
  \citenamefont {Liu}, \citenamefont {Xu}, \citenamefont {Gao}, \citenamefont
  {Zhu}, \citenamefont {Xie},\ and\ \citenamefont
  {Yang}}]{jiang2021topological}%
  \BibitemOpen
  \bibfield  {author} {\bibinfo {author} {\bibfnamefont {H.}~\bibnamefont
  {Jiang}}, \bibinfo {author} {\bibfnamefont {W.}~\bibnamefont {Liu}}, \bibinfo
  {author} {\bibfnamefont {J.}~\bibnamefont {Xu}}, \bibinfo {author}
  {\bibfnamefont {B.}~\bibnamefont {Gao}}, \bibinfo {author} {\bibfnamefont
  {C.}~\bibnamefont {Zhu}}, \bibinfo {author} {\bibfnamefont {S.}~\bibnamefont
  {Xie}},\ and\ \bibinfo {author} {\bibfnamefont {Y.}~\bibnamefont {Yang}},\
  }\bibfield  {title} {\bibinfo {title} {Topological edge modes in
  one-dimensional photonic crystals containing metal},\ }\href
  {https://doi.org/10.1364/osac.416906} {\bibfield  {journal} {\bibinfo
  {journal} {OSA Continuum}\ }\textbf {\bibinfo {volume} {4}},\ \bibinfo
  {pages} {1626} (\bibinfo {year} {2021})}\BibitemShut {NoStop}%
\bibitem [{\citenamefont {Cheng}\ \emph {et~al.}(2022)\citenamefont {Cheng},
  \citenamefont {Wang}, \citenamefont {Lv},\ and\ \citenamefont
  {Liu}}]{cheng2022topological}%
  \BibitemOpen
  \bibfield  {author} {\bibinfo {author} {\bibfnamefont {Q.}~\bibnamefont
  {Cheng}}, \bibinfo {author} {\bibfnamefont {S.}~\bibnamefont {Wang}},
  \bibinfo {author} {\bibfnamefont {J.}~\bibnamefont {Lv}},\ and\ \bibinfo
  {author} {\bibfnamefont {N.}~\bibnamefont {Liu}},\ }\bibfield  {title}
  {\bibinfo {title} {Topological photonic crystal biosensor with valley edge
  modes based on a silicon-on-insulator slab},\ }\href
  {https://doi.org/10.1364/oe.443907} {\bibfield  {journal} {\bibinfo
  {journal} {Optics Express}\ }\textbf {\bibinfo {volume} {30}},\ \bibinfo
  {pages} {10792} (\bibinfo {year} {2022})}\BibitemShut {NoStop}%
\bibitem [{\citenamefont {Imhof}\ \emph {et~al.}(2018)\citenamefont {Imhof},
  \citenamefont {Berger}, \citenamefont {Bayer}, \citenamefont {Brehm},
  \citenamefont {Molenkamp}, \citenamefont {Kiessling}, \citenamefont
  {Schindler}, \citenamefont {Lee}, \citenamefont {Greiter}, \citenamefont
  {Neupert},\ and\ \citenamefont {Thomale}}]{imhof2018topolectrical}%
  \BibitemOpen
  \bibfield  {author} {\bibinfo {author} {\bibfnamefont {S.}~\bibnamefont
  {Imhof}}, \bibinfo {author} {\bibfnamefont {C.}~\bibnamefont {Berger}},
  \bibinfo {author} {\bibfnamefont {F.}~\bibnamefont {Bayer}}, \bibinfo
  {author} {\bibfnamefont {J.}~\bibnamefont {Brehm}}, \bibinfo {author}
  {\bibfnamefont {L.~W.}\ \bibnamefont {Molenkamp}}, \bibinfo {author}
  {\bibfnamefont {T.}~\bibnamefont {Kiessling}}, \bibinfo {author}
  {\bibfnamefont {F.}~\bibnamefont {Schindler}}, \bibinfo {author}
  {\bibfnamefont {C.~H.}\ \bibnamefont {Lee}}, \bibinfo {author} {\bibfnamefont
  {M.}~\bibnamefont {Greiter}}, \bibinfo {author} {\bibfnamefont
  {T.}~\bibnamefont {Neupert}},\ and\ \bibinfo {author} {\bibfnamefont
  {R.}~\bibnamefont {Thomale}},\ }\bibfield  {title} {\bibinfo {title}
  {Topolectrical-circuit realization of topological corner modes},\ }\href
  {https://doi.org/10.1038/s41567-018-0246-1} {\bibfield  {journal} {\bibinfo
  {journal} {Nature Physics}\ }\textbf {\bibinfo {volume} {14}},\ \bibinfo
  {pages} {925–929} (\bibinfo {year} {2018})}\BibitemShut {NoStop}%
\bibitem [{\citenamefont {Liu}\ \emph {et~al.}(2022)\citenamefont {Liu},
  \citenamefont {Cao}, \citenamefont {Chen}, \citenamefont {Wang},
  \citenamefont {Yang},\ and\ \citenamefont {Zhang}}]{liu2022fully}%
  \BibitemOpen
  \bibfield  {author} {\bibinfo {author} {\bibfnamefont {Y.}~\bibnamefont
  {Liu}}, \bibinfo {author} {\bibfnamefont {W.}~\bibnamefont {Cao}}, \bibinfo
  {author} {\bibfnamefont {W.}~\bibnamefont {Chen}}, \bibinfo {author}
  {\bibfnamefont {H.}~\bibnamefont {Wang}}, \bibinfo {author} {\bibfnamefont
  {L.}~\bibnamefont {Yang}},\ and\ \bibinfo {author} {\bibfnamefont
  {X.}~\bibnamefont {Zhang}},\ }\bibfield  {title} {\bibinfo {title} {Fully
  integrated topological electronics},\ }\bibfield  {journal} {\bibinfo
  {journal} {Scientific Reports}\ }\textbf {\bibinfo {volume} {12}},\ \href
  {https://doi.org/10.1038/s41598-022-17010-8} {10.1038/s41598-022-17010-8}
  (\bibinfo {year} {2022})\BibitemShut {NoStop}%
\bibitem [{\citenamefont {Giuliani}\ and\ \citenamefont
  {Vignale}(2005)}]{giuliani2005quantum}%
  \BibitemOpen
  \bibfield  {author} {\bibinfo {author} {\bibfnamefont {G.~F.}\ \bibnamefont
  {Giuliani}}\ and\ \bibinfo {author} {\bibfnamefont {G.}~\bibnamefont
  {Vignale}},\ }\href {https://doi.org/10.1017/CBO9780511619915} {\emph
  {\bibinfo {title} {Quantum Theory of the Electron Liquid}}}\ (\bibinfo
  {publisher} {Cambridge University Press},\ \bibinfo {year}
  {2005})\BibitemShut {NoStop}%
\bibitem [{\citenamefont {Guan}\ \emph {et~al.}(2023)\citenamefont {Guan},
  \citenamefont {Haas}, \citenamefont {Schl\"omer},\ and\ \citenamefont
  {Jiang}}]{guan2022plasmons}%
  \BibitemOpen
  \bibfield  {author} {\bibinfo {author} {\bibfnamefont {Y.}~\bibnamefont
  {Guan}}, \bibinfo {author} {\bibfnamefont {S.}~\bibnamefont {Haas}}, \bibinfo
  {author} {\bibfnamefont {H.}~\bibnamefont {Schl\"omer}},\ and\ \bibinfo
  {author} {\bibfnamefont {Z.}~\bibnamefont {Jiang}},\ }\bibfield  {title}
  {\bibinfo {title} {Plasmons in ${\mathbb{z}}_{2}$ topological insulators},\
  }\href {https://doi.org/10.1103/PhysRevB.107.155414} {\bibfield  {journal}
  {\bibinfo  {journal} {Phys. Rev. B}\ }\textbf {\bibinfo {volume} {107}},\
  \bibinfo {pages} {155414} (\bibinfo {year} {2023})}\BibitemShut {NoStop}%
\bibitem [{\citenamefont {Jalali-mola}\ \emph {et~al.}(2023)\citenamefont
  {Jalali-mola}, \citenamefont {Grass}, \citenamefont {Kasper}, \citenamefont
  {Lewenstein},\ and\ \citenamefont {Bhattacharya}}]{Jalali-mola_PRL2023}%
  \BibitemOpen
  \bibfield  {author} {\bibinfo {author} {\bibfnamefont {Z.}~\bibnamefont
  {Jalali-mola}}, \bibinfo {author} {\bibfnamefont {T.}~\bibnamefont {Grass}},
  \bibinfo {author} {\bibfnamefont {V.}~\bibnamefont {Kasper}}, \bibinfo
  {author} {\bibfnamefont {M.}~\bibnamefont {Lewenstein}},\ and\ \bibinfo
  {author} {\bibfnamefont {U.}~\bibnamefont {Bhattacharya}},\ }\bibfield
  {title} {\bibinfo {title} {Topological bogoliubov quasiparticles from
  bose-einstein condensate in a flat band system},\ }\href
  {https://doi.org/10.1103/PhysRevLett.131.226601} {\bibfield  {journal}
  {\bibinfo  {journal} {Phys. Rev. Lett.}\ }\textbf {\bibinfo {volume} {131}},\
  \bibinfo {pages} {226601} (\bibinfo {year} {2023})}\BibitemShut {NoStop}%
\bibitem [{\citenamefont {Tesfaye}\ and\ \citenamefont
  {Eckardt}(2024)}]{Tesfaye_arxiv2024}%
  \BibitemOpen
  \bibfield  {author} {\bibinfo {author} {\bibfnamefont {I.}~\bibnamefont
  {Tesfaye}}\ and\ \bibinfo {author} {\bibfnamefont {A.}~\bibnamefont
  {Eckardt}},\ }\href {https://arxiv.org/abs/2406.12981} {\bibinfo {title}
  {Quantum geometry of bosonic bogoliubov quasiparticles}} (\bibinfo {year}
  {2024}),\ \Eprint {https://arxiv.org/abs/2406.12981} {arXiv:2406.12981
  [cond-mat.quant-gas]} \BibitemShut {NoStop}%
\bibitem [{\citenamefont {Lewenstein}\ \emph {et~al.}(2016)\citenamefont
  {Lewenstein}, \citenamefont {Sanpera},\ and\ \citenamefont
  {Ahufinger}}]{Lewenstein_Book2016}%
  \BibitemOpen
  \bibfield  {author} {\bibinfo {author} {\bibfnamefont {M.}~\bibnamefont
  {Lewenstein}}, \bibinfo {author} {\bibfnamefont {A.}~\bibnamefont
  {Sanpera}},\ and\ \bibinfo {author} {\bibfnamefont {V.}~\bibnamefont
  {Ahufinger}},\ }\href@noop {} {\emph {\bibinfo {title} {Ultracold Atoms in
  Optical Lattices}}}\ (\bibinfo  {publisher} {Oxford University Press},\
  \bibinfo {address} {London, England},\ \bibinfo {year} {2016})\BibitemShut
  {NoStop}%
\bibitem [{\citenamefont {Pitaevskii}\ and\ \citenamefont
  {Stringari}(2016)}]{Pitaevskii_Book2016}%
  \BibitemOpen
  \bibfield  {author} {\bibinfo {author} {\bibfnamefont {L.~P.}\ \bibnamefont
  {Pitaevskii}}\ and\ \bibinfo {author} {\bibfnamefont {S.}~\bibnamefont
  {Stringari}},\ }\href@noop {} {\emph {\bibinfo {title} {{Bose-Einstein}
  Condensation and Superfluidity}}},\ International Series of Monographs on
  Physics\ (\bibinfo  {publisher} {Oxford University Press},\ \bibinfo
  {address} {London, England},\ \bibinfo {year} {2016})\BibitemShut {NoStop}%
\bibitem [{\citenamefont {Su}\ \emph {et~al.}(1979)\citenamefont {Su},
  \citenamefont {Schrieffer},\ and\ \citenamefont {Heeger}}]{Su_PRL1979}%
  \BibitemOpen
  \bibfield  {author} {\bibinfo {author} {\bibfnamefont {W.~P.}\ \bibnamefont
  {Su}}, \bibinfo {author} {\bibfnamefont {J.~R.}\ \bibnamefont {Schrieffer}},\
  and\ \bibinfo {author} {\bibfnamefont {A.~J.}\ \bibnamefont {Heeger}},\
  }\bibfield  {title} {\bibinfo {title} {Solitons in polyacetylene},\ }\href
  {https://doi.org/10.1103/PhysRevLett.42.1698} {\bibfield  {journal} {\bibinfo
   {journal} {Phys. Rev. Lett.}\ }\textbf {\bibinfo {volume} {42}},\ \bibinfo
  {pages} {1698} (\bibinfo {year} {1979})}\BibitemShut {NoStop}%
\bibitem [{\citenamefont {Polini}\ and\ \citenamefont
  {Tosi}(2006)}]{tosi2006manybody}%
  \BibitemOpen
  \bibfield  {author} {\bibinfo {author} {\bibfnamefont {M.}~\bibnamefont
  {Polini}}\ and\ \bibinfo {author} {\bibfnamefont {M.}~\bibnamefont {Tosi}},\
  }\href@noop {} {\emph {\bibinfo {title} {Many-Body Physics in Condensed
  Matter Systems}}}\ (\bibinfo  {publisher} {Edizione Della Normale},\ \bibinfo
  {year} {2006})\BibitemShut {NoStop}%
\bibitem [{\citenamefont {Moudgil}\ \emph {et~al.}(1997)\citenamefont
  {Moudgil}, \citenamefont {Ahluwalia},\ and\ \citenamefont
  {Pathak}}]{moudgil1997ground}%
  \BibitemOpen
  \bibfield  {author} {\bibinfo {author} {\bibfnamefont {R.~K.}\ \bibnamefont
  {Moudgil}}, \bibinfo {author} {\bibfnamefont {P.~K.}\ \bibnamefont
  {Ahluwalia}},\ and\ \bibinfo {author} {\bibfnamefont {K.~N.}\ \bibnamefont
  {Pathak}},\ }\bibfield  {title} {\bibinfo {title} {Ground state of a
  double-layer charged bose system},\ }\href
  {https://doi.org/10.1103/PhysRevB.56.14776} {\bibfield  {journal} {\bibinfo
  {journal} {Phys. Rev. B}\ }\textbf {\bibinfo {volume} {56}},\ \bibinfo
  {pages} {14776} (\bibinfo {year} {1997})}\BibitemShut {NoStop}%
\bibitem [{\citenamefont {Akaturk}\ \emph {et~al.}(2018)\citenamefont
  {Akaturk}, \citenamefont {Abedinpour},\ and\ \citenamefont
  {Tanatar}}]{Akaturk_JPC2018}%
  \BibitemOpen
  \bibfield  {author} {\bibinfo {author} {\bibfnamefont {E.}~\bibnamefont
  {Akaturk}}, \bibinfo {author} {\bibfnamefont {S.~H.}\ \bibnamefont
  {Abedinpour}},\ and\ \bibinfo {author} {\bibfnamefont {B.}~\bibnamefont
  {Tanatar}},\ }\bibfield  {title} {\bibinfo {title} {Density-wave instability
  and collective modes in a bilayer system of antiparallel dipoles},\ }\href
  {https://doi.org/10.1088/2399-6528/aa9fc1} {\bibfield  {journal} {\bibinfo
  {journal} {Journal of Physics Communications}\ }\textbf {\bibinfo {volume}
  {2}},\ \bibinfo {pages} {015018} (\bibinfo {year} {2018})}\BibitemShut
  {NoStop}%
\bibitem [{\citenamefont {Seydi}\ \emph {et~al.}(2020)\citenamefont {Seydi},
  \citenamefont {Abedinpour}, \citenamefont {Zillich}, \citenamefont {Asgari},\
  and\ \citenamefont {Tanatar}}]{Seydi_PRA2020}%
  \BibitemOpen
  \bibfield  {author} {\bibinfo {author} {\bibfnamefont {I.}~\bibnamefont
  {Seydi}}, \bibinfo {author} {\bibfnamefont {S.~H.}\ \bibnamefont
  {Abedinpour}}, \bibinfo {author} {\bibfnamefont {R.~E.}\ \bibnamefont
  {Zillich}}, \bibinfo {author} {\bibfnamefont {R.}~\bibnamefont {Asgari}},\
  and\ \bibinfo {author} {\bibfnamefont {B.}~\bibnamefont {Tanatar}},\
  }\bibfield  {title} {\bibinfo {title} {Rotons and bose condensation in
  rydberg-dressed bose gases},\ }\href
  {https://doi.org/10.1103/PhysRevA.101.013628} {\bibfield  {journal} {\bibinfo
   {journal} {Phys. Rev. A}\ }\textbf {\bibinfo {volume} {101}},\ \bibinfo
  {pages} {013628} (\bibinfo {year} {2020})}\BibitemShut {NoStop}%
\bibitem [{\citenamefont {Pouresmaeeli}\ \emph {et~al.}(2023)\citenamefont
  {Pouresmaeeli}, \citenamefont {Abedinpour},\ and\ \citenamefont
  {Tanatar}}]{Pouresmaeeli_2023JPhysB}%
  \BibitemOpen
  \bibfield  {author} {\bibinfo {author} {\bibfnamefont {F.}~\bibnamefont
  {Pouresmaeeli}}, \bibinfo {author} {\bibfnamefont {S.~H.}\ \bibnamefont
  {Abedinpour}},\ and\ \bibinfo {author} {\bibfnamefont {B.}~\bibnamefont
  {Tanatar}},\ }\bibfield  {title} {\bibinfo {title} {Density and pseudo-spin
  rotons in a bilayer of soft-core bosons},\ }\href
  {https://doi.org/10.1088/1361-6455/acd599} {\bibfield  {journal} {\bibinfo
  {journal} {Journal of Physics B: Atomic, Molecular and Optical Physics}\
  }\textbf {\bibinfo {volume} {56}},\ \bibinfo {pages} {125001} (\bibinfo
  {year} {2023})}\BibitemShut {NoStop}%
\bibitem [{sup()}]{supplemental}%
  \BibitemOpen
  \href@noop {} {}\bibinfo {note} {See the Supplemental Materials for
  details.}\BibitemShut {Stop}%
\bibitem [{\citenamefont {Abramowitz}\ and\ \citenamefont
  {Stegun}(1965)}]{Abramowitz_Book1965}%
  \BibitemOpen
  \bibinfo {editor} {\bibfnamefont {M.}~\bibnamefont {Abramowitz}}\ and\
  \bibinfo {editor} {\bibfnamefont {I.~A.}\ \bibnamefont {Stegun}},\ eds.,\
  \href@noop {} {\emph {\bibinfo {title} {Handbook of mathematical
  functions}}},\ Dover Books on Mathematics\ (\bibinfo  {publisher} {Dover
  Publications},\ \bibinfo {address} {Mineola, NY},\ \bibinfo {year}
  {1965})\BibitemShut {NoStop}%
\bibitem [{\citenamefont {Chen}\ \emph {et~al.}(2024)\citenamefont {Chen},
  \citenamefont {Huang}, \citenamefont {Velkovsky}, \citenamefont {Ozawa},
  \citenamefont {Price}, \citenamefont {Covey},\ and\ \citenamefont
  {Gadway}}]{chen_arxiv2024}%
  \BibitemOpen
  \bibfield  {author} {\bibinfo {author} {\bibfnamefont {T.}~\bibnamefont
  {Chen}}, \bibinfo {author} {\bibfnamefont {C.}~\bibnamefont {Huang}},
  \bibinfo {author} {\bibfnamefont {I.}~\bibnamefont {Velkovsky}}, \bibinfo
  {author} {\bibfnamefont {T.}~\bibnamefont {Ozawa}}, \bibinfo {author}
  {\bibfnamefont {H.}~\bibnamefont {Price}}, \bibinfo {author} {\bibfnamefont
  {J.~P.}\ \bibnamefont {Covey}},\ and\ \bibinfo {author} {\bibfnamefont
  {B.}~\bibnamefont {Gadway}},\ }\href {https://arxiv.org/abs/2404.00737}
  {\bibinfo {title} {Interaction-driven breakdown of aharonov--bohm caging in
  flat-band rydberg lattices}} (\bibinfo {year} {2024}),\ \Eprint
  {https://arxiv.org/abs/2404.00737} {arXiv:2404.00737 [cond-mat.quant-gas]}
  \BibitemShut {NoStop}%
\bibitem [{\citenamefont {Sch\"{a}fer}\ \emph {et~al.}(2020)\citenamefont
  {Sch\"{a}fer}, \citenamefont {Fukuhara}, \citenamefont {Sugawa},
  \citenamefont {Takasu},\ and\ \citenamefont {Takahashi}}]{Schfer_NRP2020}%
  \BibitemOpen
  \bibfield  {author} {\bibinfo {author} {\bibfnamefont {F.}~\bibnamefont
  {Sch\"{a}fer}}, \bibinfo {author} {\bibfnamefont {T.}~\bibnamefont
  {Fukuhara}}, \bibinfo {author} {\bibfnamefont {S.}~\bibnamefont {Sugawa}},
  \bibinfo {author} {\bibfnamefont {Y.}~\bibnamefont {Takasu}},\ and\ \bibinfo
  {author} {\bibfnamefont {Y.}~\bibnamefont {Takahashi}},\ }\bibfield  {title}
  {\bibinfo {title} {Tools for quantum simulation with ultracold atoms in
  optical lattices},\ }\href {https://doi.org/10.1038/s42254-020-0195-3}
  {\bibfield  {journal} {\bibinfo  {journal} {Nature Reviews Physics}\ }\textbf
  {\bibinfo {volume} {2}},\ \bibinfo {pages} {411–425} (\bibinfo {year}
  {2020})}\BibitemShut {NoStop}%
\end{thebibliography}%

\newpage
\pagebreak
\clearpage
\widetext

\begin{center}
	\textbf{\large Supplementary materials for Engineering the Bogoliubov Modes through Geometry and Interaction: From Collective Edge Modes to Flat-band Excitations}
\end{center}

\setcounter{equation}{0}
\setcounter{figure}{0}
\setcounter{table}{0}
\setcounter{page}{1}
\makeatletter
\renewcommand{\theequation}{S\arabic{equation}}
\renewcommand{\thefigure}{S\arabic{figure}}
\renewcommand{\bibnumfmt}[1]{[S#1]}
\renewcommand{\citenumfont}[1]{S#1}

\section{Fourier transforms of the soft-core interactions}
	In the SSH model of BECs, if we consider a soft-core and short-range interaction between particles, modeled with a step-function potential, the same tube (S), intra-cell (I), and inter-cell (O) interactions in the real space read
	\be
	\begin{split}
		{u_{{\mathrm{S}}}}(r) =& U\Theta ({r_c} - r), \\
		{u_{{\mathrm{I}}}}(r) =& U\Theta ({r_c} - \sqrt {{r^2} + {d^2}} ), \\
		{u_{{\mathrm{O}}}}(r) =& U\Theta ({r_c} - \sqrt {{r^2} + {(a-d)^2}} ),
	\end{split} 
	\ee
where ${U}$ and ${r_c}$ are the interaction strength and the soft-core radius, respectively, and $\Theta (x)$ is the Heaviside step function. 
The 2D Fourier transforms of these soft-core interactions are readily obtained as
	\be
	\begin{split}
		u_{\mathrm{S}}(q) =& 2\pi U r_c J_1\left(r_c q\right)/q, \\
		u_{\mathrm{I}}(q) =& 2\pi U \sqrt {{r_c}^2 - {d^2}} J_1\left(\sqrt {{r_c}^2 - {d^2}} q\right)/q,\\
		u_{\mathrm{O}}(q) =& 2\pi U \sqrt {r_c^2 - (a-d)^2} J_1\left(\sqrt {r_c^2 - (a-d)^2} q\right)/q,
	\end{split} 
	\ee 
	where ${J_1}(x)$ is the Bessel function of the first kind.

\section{Effects of second neighbor interaction in the SSH superlattice}\
If we restore the effects of the interaction between second neighbor layers $u_{\rm n}(q)$ in the Su-Schrieffer-Heeger model, the elements of the dynamical matrix in Eq.\eqref{d_ssh_gen} read
	\be\label{}
	\begin{split}
		\mathcal{D}_{A A}(q, k)&=\varepsilon_{q}^{2}+2 n \varepsilon_{q}\left[u_{\rm S}(q)+2 u_{\rm n}(q) \cos(ka)\right] , \\
		\mathcal{D}_{A B}(q, k)&= 2 n \varepsilon_{q} e^{i k d}\left[u_\mathrm{I}(q)+ e^{-i k a} u_\mathrm{O}(q)\right].
	\end{split}
	\ee
	Here, again we have $\mathcal{D}_{BB}=\mathcal{D}_{A A}$, and $\mathcal{D}_{ BA}=\mathcal{D}^*_{A B}$.
	The eigenvalues of this ${\mathcal D}(q,k)$-matrix are the square modules of the collective mode dispersions 
	\be
		E_{\pm}^{2}(q, k)=\varepsilon_{q}^{2}+2 n \varepsilon_{q}\left[u_{\rm S}(q)+2 u_{\rm n}(q) \cos (k a)\right] 
		\pm 2 n \varepsilon_{q}\sqrt{u_{\mathrm{I}}^{2}(q)+u_{\mathrm{O}}^{2}(q)+2  u_{\mathrm{I}}(q) u_{\mathrm{O}}(q)\cos (k a) }.
	\ee
	Furthermore, we can find the zero-sound velocities from the long-wavelength limits of the collective modes
	\be
		v_{\pm}(k)=\lim _{q \rightarrow 0} \frac{E_{\pm}(q, k)}{\hbar q} 
		=v_s\sqrt{1+2 {\bar u}_{\mathrm{n}} \cos (k a)\pm \sqrt{{\bar u}_{\mathrm{I}}^{2}+{\bar u}_{\mathrm{O}}^{2}+2 {\bar u}_{\mathrm{I}} {\bar u}_{\mathrm{O}} \cos (k a)}},
	\ee
	where ${\bar u}_{{\rm n}}\equiv u_{{\rm n}}(q=0)/u_{\rm S}(q=0)$.
	The effects of second neighbor interaction on the sound velocity are illustrated in the right panel of Fig.~\ref{fig:zero-sound2}.
\end{document}